\documentclass[useAMS,usenatbib]{mn2e}
\usepackage{amsmath}
\usepackage{graphicx,amssymb,amstext} 
\usepackage{float} 

\def\aj{AJ}			
		
\def\apj{ApJ}

\def\aap{A\&A}		
		
\def\aaps{A\&AS}

\def\mnras{MNRAS}

\begin{document}

\title[HerMES: Point source catalogues from \emph{Herschel}-SPIRE observations II]{HerMES: Point source catalogues from \emph{Herschel}-SPIRE observations II\thanks{Herschel is an ESA space observatory with science instruments provided by European-led Principal Investigator consortia and with important participation from NASA.}}

\author[L.~Wang et al.]
{\parbox{\textwidth}{\raggedright L.~Wang,$^{1, 2}$\thanks{E-mail: \texttt{lingyu.wang@durham.ac.uk}}
M.~Viero,$^{3}$
C. ~Clarke,$^{1}$
J.~Bock,$^{3,4}$
V.~Buat,$^{5}$
A.~Conley,$^{6}$
D.~Farrah,$^{1, 7}$
S.~Heinis,$^{5}$
G.~Magdis,$^{8}$
L.~Marchetti,$^{9, 10}$
G.~Marsden,$^{11}$
P.~Norberg,$^2$
S.J.~Oliver,$^{1}$
Y.~Roehlly,$^{5}$
I.G.~Roseboom,$^{1,12}$
B.~Schulz,$^{3,13}$ 
A.J.~Smith,$^{1}$
M.~Vaccari,$^{9, 14}$ and
M.~Zemcov$^{3,4}$}\vspace{0.4cm}\\
\parbox{\textwidth}{\raggedright $^{1}$Astronomy Centre, Dept. of Physics \& Astronomy, University of Sussex, Brighton BN1 9QH, UK\\
$^{2}$Institute for Computational Cosmology, Department of Physics, Durham University, South Road, Durham DH1 3LE, UK\\
$^{3}$California Institute of Technology, 1200 E. California Blvd., Pasadena, CA 91125, USA\\
$^{4}$Jet Propulsion Laboratory, 4800 Oak Grove Drive, Pasadena, CA 91109, USA\\
$^{5}$Laboratoire d'Astrophysique de Marseille, OAMP, Universit\'e Aix-marseille, CNRS, 38 rue Fr\'ed\'eric Joliot-Curie, 13388 Marseille cedex 13, France\\
$^{6}$Center for Astrophysics and Space Astronomy 389-UCB, University of Colorado, Boulder, CO 80309, USA\\
$^{7}$Department of Physics, Virginia Tech, Blacksburg, VA 24061, USA\\
$^{8}$Denys Wilkinson Building, Keble Road, University of Oxford, Oxford, OX1 3RH, UK\\
$^{9}$Dipartimento di Astronomia, Universit\`{a} di Padova, vicolo Osservatorio, 3, 35122 Padova, Italy\\
$^{10}$Department of Physical Sciences, The Open University, Milton Keynes MK7 6AA, UK\\
$^{11}$Department of Physics \& Astronomy, University of British Columbia, 6224 Agricultural Road, Vancouver, BC V6T~1Z1, Canada\\
$^{12}$Institute for Astronomy, University of Edinburgh, Royal Observatory, Blackford Hill, Edinburgh EH9 3HJ, UK\\
$^{13}$Infrared Processing and Analysis Center, MS 100-22, California Institute of Technology, JPL, Pasadena, CA 91125, USA\\
$^{14}$Astrophysics Group, Physics Department, University of the Western Cape, Private Bag X17, 7535, Bellville, Cape Town, South Africa}}

\date{Accepted . Received ; in original form }

\maketitle

\begin{abstract}

The {\it Herschel} Multi-tiered Extragalactic Survey (HerMES) is the largest Guaranteed Time Key Programme on the {\it Herschel} Space Observatory. With a wedding cake survey strategy, it consists of nested fields with varying depth and area totalling $\sim$380 deg$^2$. In this paper, we present deep point source catalogues extracted from {\it Herschel}-SPIRE observations of all HerMES fields, except for the later addition of the 270 deg$^2$ HeLMS field. These catalogues constitute the second Data Release (DR2) made in October 2013. A subset of these catalogues, which consists of bright sources extracted from {\it Herschel}-SPIRE observations completed by May 1, 2010 (covering $\sim$ 74 deg$^2$) were released earlier in the first extensive Data Release (DR1) in March 2012. Two different methods are used to generate the point source catalogues, the SUSSEXtractor (SXT) point source extractor used in two earlier data releases (EDR and EDR2) and a new source detection and photometry method. The latter combines an iterative source detection algorithm, StarFinder (SF), and a De-blended SPIRE Photometry (DESPHOT) algorithm. We use end-to-end {\it Herschel}-SPIRE simulations with realistic number counts and clustering properties to characterise basic properties of the point source catalogues, such as the completeness, reliability,  photometric and positional accuracy. Over 500, 000 catalogue entries in HerMES fields (except HeLMS) are released to the public through the HeDAM website (\url{http://hedam.oamp.fr/herMES}).

\end{abstract}


\section{INTRODUCTION}

The {\it Herschel} Multi-tiered Extragalactic Survey (HerMES\footnote{\url{http://hermes.sussex.ac.uk}}) (Oliver et al. 2012) is a Guaranteed Time Key Program on the {\it Herschel} Space Observatory (Pilbratt et al. 2010). It has a wedding cake survey design which consists of nested fields ranging from shallow and wide fields to deep and narrow fields observed with the {\it Herschel}-Spectral and Photometric Imaging Receiver (SPIRE; Griffin et al. 2010) at 250, 350 and 500 $\micron$\footnote{The SPIRE bands at 250, 350 and 500 $\micron$ are also known as the SPIRE Photometer Short Wavelength array (PSW), SPIRE Photometer Median Wavelength array (PMW) and SPIRE Photometer Long Wavelength array (PLW) respectively.} and the {\it Herschel}-Photodetector Array Camera and Spectrometer (PACS; Poglitsch et al. 2010) at 100 and 160 $\micron$ for a subset of the HerMES fields. There are 13 target blank fields at approximately seven different depths (Levels 1-7) covering a total area of $\sim$380 deg$^2$ which include a later addition of a wide HerMES Large-Mode Survey (HeLMS) field (270 deg$^2$) observed by SPIRE alone. In addition to the blank fields, HerMES also targeted 12 known clusters. The first two data releases, Early Data Release (EDR, July 1, 2010) and EDR2 (September 19, 2011), included SPIRE high signal-to-noise (${\rm SNR}\ge 5$) sources extracted from HerMES Science Demonstration Phase (SDP\footnote{The SDP fields include the First Look Survey (FLS), GOODS-N, Lockman-SWIRE and Lockman-North.}) and the first data release (DR1) fields (see Table 1) generated by the SUSSEXtractor (SXT) point source extractor (Smith et al. 2012) as well as SPIRE maps in the Abell 2218 cluster field.

This paper describes the generation of HerMES point source catalogues extracted from {\it Herschel}-SPIRE observations completed by May 1, 2010 and released during the first extensive Data Release (DR1) of maps and catalogues (March 27, 2012) and all observations except HeLMS released during the second extensive Data Release (DR2; October 31, 2013). Details of DR1 and DR2 are given in Table 1 and 2. As catalogues are the starting point for understanding the far-infrared / sub-millimetre (sub-mm) galaxy population in detail (e.g.  their spectral energy distributions, redshift distribution and luminosity), a lot of effort has been invested in constructing deep and reliable catalogues. The main challenge is confusion noise which arises when the spatial extent of the emission from distinct sources overlap within the same area, creating signal fluctuations within the telescope beam. At a given wavelength, this will mostly depend on the intrinsic flux density distribution of sources as well as the resolving power and sensitivity of the instrument used for the observations. Nguyen et al. (2010) found that in the limit of infinite integration time (i.e. negligible instrumental noise) the SPIRE confusion noise is at the level of $5\sigma=24.0, 27.5, 30.5$ mJy at 250, 350 and 500 $\micron$, respectively (after excluding map pixels at $>=5\sigma$). Confusion noise is a significant feature (much larger than instrumental noise) for most of the HerMES fields (from Level 1 to Level 4; see Table 1 and 2) and sets a fundamental limit on the flux limit of sources that can be detected by a peak finding algorithm such as SXT .

Bright sources that can be resolved individually by {\it Herschel} only account for a small fraction of the cosmic infrared background (CIB) (e.g. Oliver et al. 2010; Glenn et al. 2010; Bethermin et al. 2012).  To extract deeper catalogues, we must reduce the level of confusion noise in our maps. In this paper, we present a new source detection and photometry method which combines an iterative source detection algorithm StarFinder (SF) and a De-blended SPIRE Photometry (DESPHOT) algorithm. SF iteratively detects and removes sources to reduce the confusion noise level and therefore can extract sources below the nominal confusion limit. DESPHOT is optimised for accurate photometry in highly confused images.

The paper is organised as follows. In Section 2, first we describe SPIRE observations and the extracted data products of the HerMES fields released in DR1 and DR2. Then, we describe in detail the two different source extraction methods (SXT and SF combined with DESPHOT) used to generate the DR1 and DR2 point sources catalogues. In Section 3, realistic end-to-end {\it Herschel}-SPIRE simulations are used to understand the basic properties (e.g. photometric and positional error, completeness, and reliability) of the point source catalogues. The issue of extended sources being broken up by our source extraction methods is discussed in Section 4. Finally, we give conclusions and discussions in Section 5.

\section{HerMES DR1 and DR2 point source catalogues}

\begin{table*}
\caption{Summary of the HerMES observations released in DR1. The columns are the set identification number, the design level, the target name,  the observing mode, the area of good pixels $\Omega_{\rm good}$ where the number of bolometer samples per pixel in the 250 $\mu$m map is greater than half of the median value, and the $5\sigma$ instrumental noise level at 250, 350 and 500 $\micron$.}
\begin{tabular}{llllllll }
\hline
Set & Level & Target & Mode & $\Omega_{\rm good}$ (deg$^2$) & $5\sigma_{250}^{\rm ins.}$ (mJy) & $5\sigma_{350}^{\rm ins.}$ (mJy) & $5\sigma_{500}^{\rm ins.}$ (mJy)\\
\hline
  1 & CD & Abell 2218                 & Sp. Nom. & 0.10 & 6.4 & 5.3 &7.6 \\
  3 & CD & MS0451.6-0305       & Sp. Nom. & 0.08 & 9.2 & 7.7 & 11.0    \\
  7 & CS & Abell 2219                 & Sp. Nom. & 0.08 & 9.2 & 7.7 & 11.0  \\
  14 & L2 & GOODS-N                & Sp. Nom. & 0.55 & 3.8 & 3.1 &  4.5\\
  15 & L2 & ECDFS                     & Sp. Nom. & 0.58 & 4.3 & 3.6 & 5.2 \\
  17 & L3 & Groth Strip               & Sp. Nom. & 0.60 & 10.7 & 8.9 & 12.8 \\
  19 & L3 & Lockman-North       & Sp. Nom. & 0.65 & 10.6  & 8.8  & 12.7 \\
  28 & L5 & Lockman SWIRE    & Sp. Fast & 17.37 & 13.6    & 11.2 & 16.2 \\
  30 & L5 & Bootes HerMES      & Parallel & 3.25 & 13.8 & 11.3   & 16.4  \\
  31 & L5 & ELAIS N1 HerMES & Parallel & 3.25 & 13.8 & 11.3   & 16.4  \\
  36 & L6 & XMM-LSS SWIRE   & Parallel & 18.87  & 11.2 & 9.3 & 13.4 \\
  37 & L6 & Bootes NDWFS       & Parallel & 10.57 & 13.8 & 11.3 & 16.4   \\
  38 & L6 & ADFS                        & Parallel      & 7.47 & 25.8&21.2 & 30.8  \\
  40 & L6 & FLS                            & Parallel         & 6.71 & 25.8 & 21.2 &  30.8 \\
\hline
\end{tabular}
\label{tab:AORs}
\end{table*}

\begin{table*}
\caption{Summary of the additional HerMES observations released in DR2. The columns are the same as in Table 1.}
\begin{tabular}{llllllll }
\hline
Set & Level & Target & Mode & $\Omega_{\rm good}$ (deg$^2$) & $5\sigma_{250}^{\rm ins.}$ (mJy) & $5\sigma_{350}^{\rm ins.}$ (mJy) & $5\sigma_{500}^{\rm ins.}$ (mJy)
  \\
\hline
 2 & CD & Abell 1689 & Sp. Nom.  & 0.08 & 9.2 & 7.7 & 11.0\\
  4 & CS & RXJ13475-1145 & Sp. Nom.  & 0.08 & 9.2 & 7.7 & 11.0\\
  5 & CS & Abell 1835 & Sp. Nom.  & 0.08 &9.2 & 7.7 & 11.0\\
  6 & CS & Abell 2390 & Sp. Nom.  & 0.08 &9.2 & 7.7 & 11.0\\
  8 & CS & Abell 370 & Sp. Nom. &0.08 &9.2 & 7.7 & 11.0 \\
  9 & CS & MS1358+62 & Sp. Nom. & 0.08 &9.2 & 7.7 & 11.0\\
  10 & CS & Cl0024+16 & Sp. Nom.  & 0.08 &9.2 & 7.7 & 11.0\\
  11 & CH & MS1054.4-0321 & Sp. Nom.  & 0.16 & 13.9 & 11.6 & 16.7\\
  12 & CH & RXJ0152.7-1357 & Sp. Nom. &  0.16 & 13.9 & 11.6 & 16.7\\
  13 & L1 & GOODS-S & Sp. Nom. & 0.35& 4.3 & 3.6 & 5.2\\
  22 & L2 & COSMOS & Sp. Nom. & 2.82& 8.0 & 6.6 & 9.5\\
  18 & L3 & Lockman-East ROSAT & Sp. Nom. & 0.57 & 9.6 & 7.9 &11.5\\
  18B & L3 & Lockman-East Spitzer & Sp. Nom. & 1.40&9.6 & 7.9 &11.5 \\
  23 & L4 & UDS & Sp. Nom. & 2.02& 11.2 & 9.3 & 13.4\\
  24 & L4 & VVDS & Sp. Nom. &2.02& 11.2 & 9.3 & 13.4\\
 22B & L5 & COSMOS HerMES &  Sp. Nom.   &4.38& 15.9 & 13.3 & 19.1 \\
  27 & L5 & CDFS SWIRE & Sp. Fast & 11.39& 12.7 & 10.5 & 15.2 \\
  28B & L5 & Lockman SWIRE & Sp. Fast &  7.63&  13.6 & 11.2 & 16.2\\
  29 & L5 & EGS HerMES & Parallel & 2.67 & 10.7 & 8.9 & 12.8\\
  32 & L5 & XMM VIDEO1 & Parallel & 2.72& 14.9 & 12.2 & 17.8\\
  32B & L5 & XMM VIDEO2 & Parallel & 1.74& 14.9 &  12.2 & 17.8\\
  32C & L5 & XMM VIDEO3 & Parallel &2.73& 14.9 &  12.2 & 17.8\\
  33 & L5 & CDFS SWIRE & Parallel &  10.89&  8.0 & 6.6 & 9.6\\
  34 & L5 & Lockman SWIRE & Parallel & 16.08& 9.6 & 7.9 & 11.5\\
  39B & L5 & ELAIS S1 VIDEO & Parallel & 3.72& 14.9& 12.2 & 17.8\\
  35 & L6 & ELAIS N1 SWIRE & Parallel & 12.28 & 25.8 & 21.2 & 30.8\\
  39 & L6 & ELAIS S1 SWIRE & Parallel & 7.86&25.8 & 21.2 & 30.8 \\
  41 & L6 & ELAIS N2 SWIRE & Parallel & 7.80& 25.8 & 21.2 & 30.8\\
  
\hline
\end{tabular}
\label{tab:AORs}
\end{table*}

\subsection{Overview of DR1 and DR2 SPIRE observations and data products}

HerMES DR1 includes bright sources (above 55, 55, and 30 mJy at 250, 350 and 500 $\micron$ respectively) extracted from the SDP observations as well as all SPIRE observations completed by 2010 May 1. Table 1 gives a summary of the HerMES observations released in DR1, including the set identification number\footnote{The set identification number is defined in Oliver et al. (2012). Observations of the same field at the same level made with the same mode and areal size are grouped into a `set' .}
, the design level, the target name, the observing mode (including the nominal SPIRE scan rate at 30\arcsec\ s$^{-1}$, the fast SPIRE scan rate at 60\arcsec\ s$^{-1}$ and the SPIRE-PACS parallel mode), the area of good pixels $\Omega_{\rm good}$ where the number of bolometer samples per pixel in the 250$\,$\micron\ map is greater than half of the median value, and the $5\sigma$ instrumental noise level at 250, 350 and 500 $\micron$. HerMES DR2 includes all point sources from the SDP and  DR1 fields as well as all subsequent SPIRE observations except HeLMS. Table 2 gives a summary of the additional HerMES fields included in DR2.

We provide three different types of point source catalogues extracted from the SMAP\footnote{HerMES maps created by a combination of standard ESA software and a customised software package SMAP. For full details of the SMAP map-making pipeline, please refer to Levenson et al. (2010) and Viero et al. (2013).} v4.1 maps:

\begin{itemize}
\item Independent single-band SUSSEXtractor (SXT) catalogues at 250, 350 and 500 $\micron$. SXT is used to detect point sources and estimate their positions and fluxes.

\item Independent single-band StarFinder (SF) catalogues with DESPHOT photometry  at 250, 350 and 500 $\micron$. SF is used to detect sources and find their optimal positions, while DESPHOT is used to estimate fluxes for a given list of source positions. For convenience, we will refer to these catalogues as SF catalogues.

\item Band-merged SF catalogues with DESPHOT multi-band (250, 350 and 500 $\micron$) photometry at the positions of the SF 250 $\mu$m sources. We will refer to these catalogues as SF250 catalogues.

\end{itemize}

These point source catalogues can be downloaded from the Herschel Database in Marseille (HeDaM; \url{http://hedam.oamp.fr/herMES}). Apart from the SF250 catalogues, sources are directly detected in the image where we want to perform photometry, with no additional information obtained from other wavelengths. As a result, it allows detection of sources which might be unidentified at other wavelengths. However, when source density is too high, blind source extraction can not separate blended point sources. The source centroid from blind source catalogues might be less well constrained causing greater difficulty in cross-matching sources detected at different wavelengths. Admittedly, source extraction with prior information (e.g., from deep 24 \micron\ observations) on the spatial distribution of sources in the sky (Roseboom et al. 2010, 2012) will in general provide deeper catalogues and more robust source identification across different wavelengths. But it can risk misidentifying sources with positive noise fluctuations, if we assume that all sources in the prior model have a counterpart in the SPIRE maps.

\subsection{SUSSEXtractor (SXT) vs StarFinder (SF)}

SXT is a peak finding algorithm (implemented in IDL and Java within HIPE\footnote{The software package for Herschel Interactive Processing Environment (HIPE) is the application that allows users to work with the Herschel data, including finding the data products, interactive analysis, plotting of data, and data manipulation.}) optimised for isolated sources. For more details on the SXT source extraction method, please refer to Savage \& Oliver (2007) and Smith et al. (2011). SF is an iterative source finding and fitting program (implemented in IDL), originally designed for crowded stellar fields analysis (Diolaiti et al. 2000). SF, thus, is expected to do better at de-blending sources and finding faint sources around bright sources. SF models the observed image as a superposition of shifted scaled replicas of the Point Response Function (PRF) lying on a smooth background. At each iteration, SF performs the following steps:
\begin{enumerate}
\item Detects new sources by searching for local maxima above a given SNR threshold in the image after subtraction of the known sources.
\item Cross-correlates each of the sub-images centred at the newly detected sources with the PRF and accepts those with correlation coefficients (a measure of the similarity between the source profile and our template) above a given threshold. 
\item For each of the accepted new sources, determines the best-fit position and flux of the source of interest by fitting to the sub-image centred around the source. Adds all of the new sources with the optimal positions and fluxes to the list of accepted sources and repeats from step (i) with a lower SNR threshold.
\end{enumerate}

In both source extraction methods, for computational efficiency we use a Gaussian shaped PRF with the full width half maximum (FWHM) set to 18.15\arcsec, 25.15\arcsec\ and 36.3\arcsec\ at 250, 350 and 500 $\micron$ respectively (although the SPIRE beams are known to be significantly elliptical). As our source photometry is derived from profile fitting, aperture correction is not needed. In principle, we could use a more realistic PRF such as the beam measured from maps of Neptune (a strong point-like source). However, we find that the Gaussian PRF is very good approximation of the real PRF and there is no bias in the flux density measurement for bright sources (see Section 3.3).  For faint sources, confusion noise and instrument noise cause a systematic overestimation of the source flux (flux boosting) which is  much larger than the photometric uncertainty caused by the Gaussian approximation of the PRF.

In principle, SF should extract a deeper source catalogue than SXT and return more accurate source positions and fluxes. However, when the instrument noise level is high (e.g. our Level 5 and Level 6 observations) or when the source profile is not well sampled, fewer sources would pass the correlation test and therefore SF would return a shallower catalogue than SXT.

\subsection{The De-blended SPIRE Photometry (DESPHOT) algorithm}

While SF is effective at identifying ``peaks'' in crowded images, it is not optimised for accurate photometry in highly confused images such as those from {\it Herschel}-SPIRE. The primary reason is that it requires a large fraction of ``sky'' pixels which are free from any source flux. Having a large number of ``sky'' pixels allows the background to be accurately determined, and also for crowded clumps of sources to be isolated, i.e. sources may be blended together but are typically separated from other sources enough that placing a annulus  around them is appropriate. In {\it Herschel}-SPIRE images, nearly every pixel is dominated by signal from sources, meaning that the background actually comes from blended sources which we are trying to extract. 

To deal with these issues we have developed a new algorithm for SPIRE source photometry, {\sc DESPHOT} (De-blended SPIRE Photometry). Many of the details of the algorithm have been presented in Roseboom et al.\ (2010; 2012) in the context of cross-identifications with 24 $\mu$m and radio sources. However a complete description is provided here for the sake of clarity. {\sc DESPHOT} consists of the following conceptually distinct steps: map segmentation, source photometry, background estimation and noise estimation. We will explain each step in turn.

While in theory source photometry and background estimation do not require segmentation of the map, in practice it is often computationally infeasible to use the full image. We need to  break the map into smaller segments that can then be processed independently without affecting the photometric accuracy. This is achieved by locating islands of high SNR pixels enclosed by low SNR pixels. The segmentation algorithm operates thus:
\begin{enumerate}
\item Locates all pixels with a SNR above some threshold (default value of SNR$=1$);
\item Takes the first of these high SNR pixel starting in the bottom left corner of the image;
\item ``Grows'' a region around this pixel by iteratively taking neighbouring high SNR pixels;
\item Once there are no more high SNR neighbours jumps to the next high SNR pixel and repeat from step (iii).
\end{enumerate}
Each of these independent regions of high SNR pixels is uniquely identified and will be processed separately by the source photometry component.

{\sc DESPHOT} assumes that the map can be described as the sum of the flux densities from the $n$ known sources
\begin{equation}
{\bf d}=\sum_{i=1}^n{\bf P_i} f_i+\boldsymbol \delta, 
\end{equation}
where {\bf d} is the image, ${\bf P_i}$ the PRF for source $i$, {\bf $f_i$} the flux density of source $i$ and ${\boldsymbol  \delta}$ some unknown noise term. A linear equation of this form will have a maximum likelihood solution of the form;
\begin{equation}
{\bf \hat{f}}=({\bf A}^{\rm T}{\bf N_d}^{-1}{\bf A})^{-1}\, {\bf A}^{\rm T}{\bf N_d}^{-1}{\bf d}
\label{eqn:mlflux}
\end{equation}
where {\bf A} is a $m$ pixel by $n$ source matrix which describes the PRF for each source and ${\bf N_d}=\left<\boldsymbol \delta \boldsymbol  \delta^T\right>$ is the covariance matrix between the image pixels (assumed to be diagonal here). This equation can be solved directly, either by brute-force matrix inversion or via other linear methods (e.g. conjugate gradient methods) but this class of solution can create two significant problems. First, it ignores our prior knowledge that sources can not have negative flux density. This is not just a conceptual annoyance, as in very degenerate cases (i.e. two sources very close together) the lack of a non-negative prior results in any symmetric pairing of positive and negative flux providing a good fit to the data. The second issue is over-fitting. If we cannot provide a 100\% reliable input list, a simple linear solution does not have the power to discriminate between spurious and real sources and can result in the overall flux densities of real sources to be underestimated (as some flux is lost to spurious ones).

To solve both issues, Roseboom et al.\ (2012) introduced the non-negative, weighted, LASSO algorithm (Tibshirani 1996; Zou 2006; ter Braak et al\ 2010). LASSO belongs to a class of methods known as ``active'' set, in that it considers the solution vector (in this case the flux densities of the sources) to be either ``active'', and to be optimised in the solution, or ``inactive'', and set to zero. Basically the algorithm is iterative. It starts with the solution flux vector set to zero. It then turns on a single source at a time (i.e. moves them to the active set) that has the largest partial derivative of the chi-squared $\chi^2$ (i.e. the source that reduces the chi-squared the most). The non-negative prior means that the derivative is only considered for positive values of the flux, and the step taken in each iteration is the largest possible that keeps  all the sources in the active set positive, and the activated source $d\chi^2/df$ negative. This process continues until some tolerance level is reached. The active set approach allows the source photometry algorithm to remove sources which are not necessary to provide a good fit to the map, thus alleviating concerns about over-fitting.

Next, we need to be able to estimate the level of background emission. SPIRE does not measure the absolute background level. As a result, the background level is unknown and all Herschel-SPIRE maps have been mean subtracted, i.e. the mean of the map is zero. In reality the background emission mostly comes from real sources (not in our detection list) so they would be correlated. For simplicity, we will model it as a constant background, and solve for this iteratively starting with the assumption that it is zero.

In {\sc DESPHOT}, LASSO is used to solve Eq.~\ref{eqn:mlflux} for each segment assuming no background. Then, the background, assumed to have a fixed value across the whole map, is estimated using the first-pass photometry values. Thus our model for the map is actually;
\begin{equation}
{\bf d}=\sum_i^n{\bf P_i} f_i+B+\boldsymbol  \delta,
\end{equation}
where $B$ is the fixed background. While a solution for the background could have been incorporated into Eq.~\ref{eqn:mlflux}, because we treat each map segment independently in the source photometry step it would not be possible to produce a single value for the entire map in this way. However given a set of initial estimates for the fluxes $f_i^0$ we can estimate a value for the background $B$ via
\begin{equation}
B={\bf d}-\sum_i^n{\bf P_i} f_i^0.
\end{equation}
Once the background value is established the source photometry process is run again with the background subtracted. The flux density estimates from this second pass are the ones which enter the output catalogues.

Finally, we estimate the total noise on our sources, including the effect of confusion. As the use of LASSO and non-negative prior make the source photometry method non-linear, the most obvious way to estimate the noise would be a Monte-Carlo simulation of the full {\sc DESPHOT} algorithm. However, given that the typical processing time for a L5 field is $\sim2$ days on a large supercomputing node ($\sim20$ cores and 256\,GB of RAM) running simulations on this scale is not feasible. If we approximate the {\sc DESPHOT} algorithm as linear then we can get a lower limit on the noise via
\begin{equation}
{\bf N_f}\ge{({\bf A^TN_d^{-1}A})}^{-1}.
\end{equation}
However this estimate only includes the instrumental noise (via ${\bf N_d}$) and the degeneracies between the input sources (via {\bf A}). In order to provide some estimate of the remaining confusion noise (i.e. the contaminating fluxes from sources not in the input) we use the global map statistics via the pixel intensity distribution. Specifically, we produce a single estimate of the residual confusion noise by measuring the standard deviation of pixels in the residual map, i.e. the SPIRE map with a reconstructed model using our final estimates of the background and source fluxes removed. The instrumental noise  in the residual map must be removed to produce a clean estimate of the confusion noise. Thus the confusion noise $\sigma_{\rm{conf}}$ is calculated by taking the standard deviation of the residual map pixels $\sigma_{{\rm res}}$ and removing the average instrumental noise in these pixels in quadrature, $\sigma_{{\rm conf}}^2=\sigma_{{\rm res}}^2-\sigma_{{\rm pix}}^2$, where $\sigma_{{\rm pix}}$ is calculated directly from the exposure time per pixel. The total noise $\sigma_{{\rm tot}}$ for a point source is then calculated from both the instrumental noise (and confusion noise from the known sources), $\sigma_{i}=\sqrt{{\rm diag((A^TN_d^{-1}A})^{-1})}$, and confusion noise from the unknown sources in the residual map $\sigma_{{\rm conf}}$ via $\sigma_{{\rm tot}}^2=\sigma_i^2+\sigma_{\rm conf}^2.$

\section{Properties of the DR1 and DR2 point source catalogues}

In this section, we will discuss the properties of HerMES  DR1 and DR2 SXT, SF and SF250 catalogues with realistic end-to-end simulations designed to match real {\it Herschel}-SPIRE observations as well as the map-making and the point source extraction process.

\begin{figure}
\centering
\includegraphics[height=2.6in,width=3.4in]{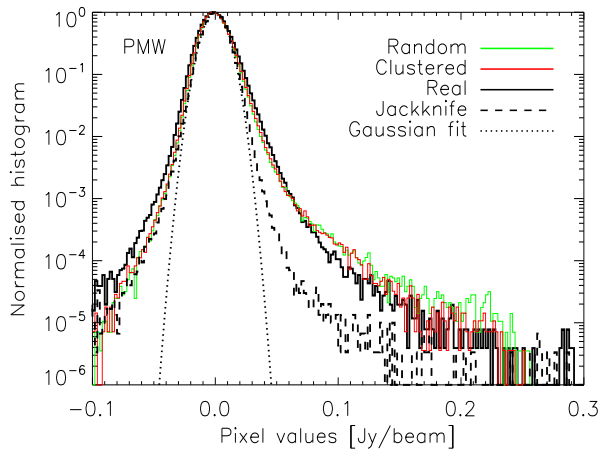}
\caption{Histogram of pixel flux densities of the real map  (black solid line), simulated maps (green line: random positions; red line: clustered positions), and jackknife noise map (black dashed line) of the Lockman-SWIRE field at 350 $\micron$ (PMW). The black dotted line is a Gaussian fit to the pixel histogram of the jackknife noise map.}
\label{pofd}
\end{figure}

\begin{figure*}
\centering
\includegraphics[width=2.3in]{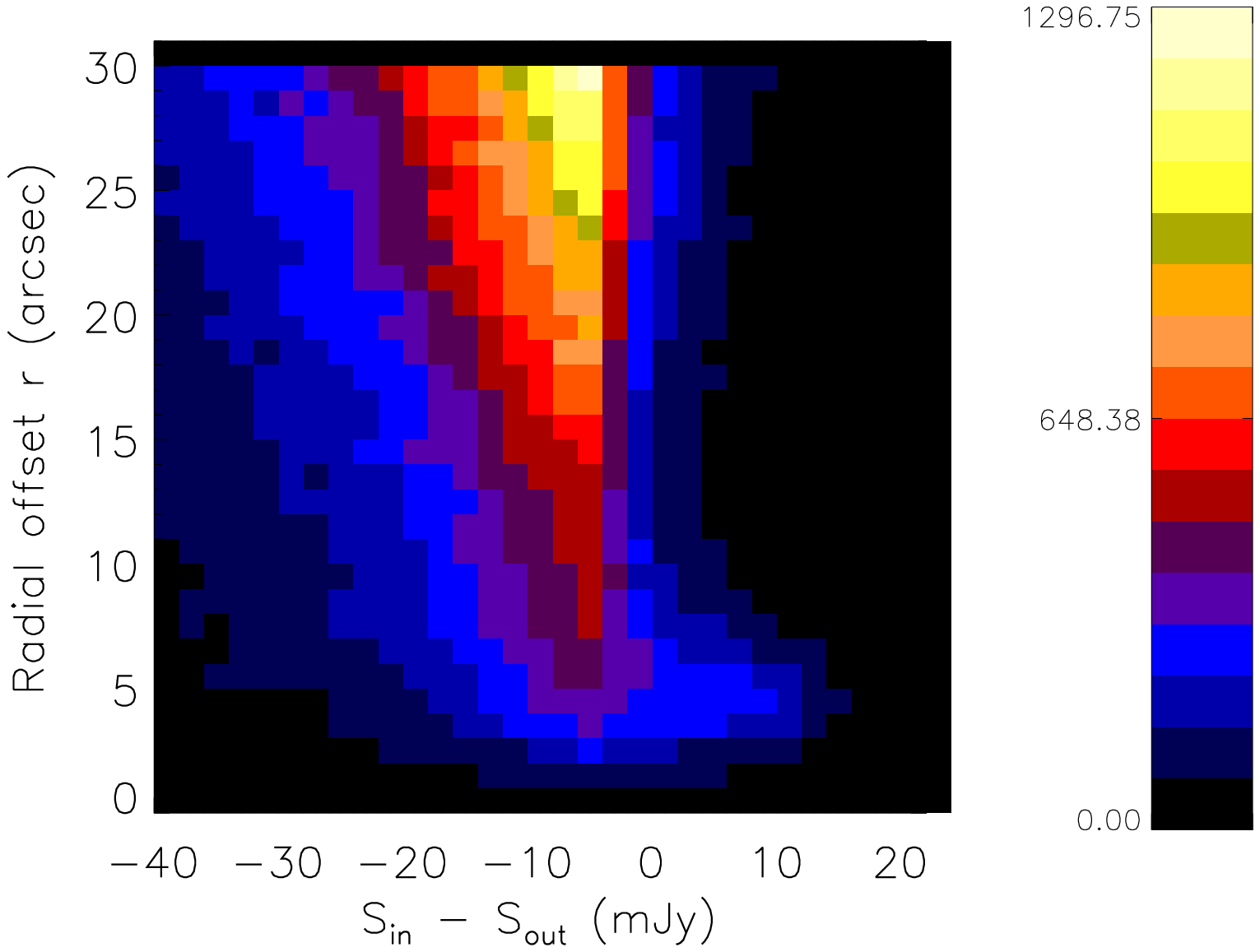}
\includegraphics[width=2.3in]{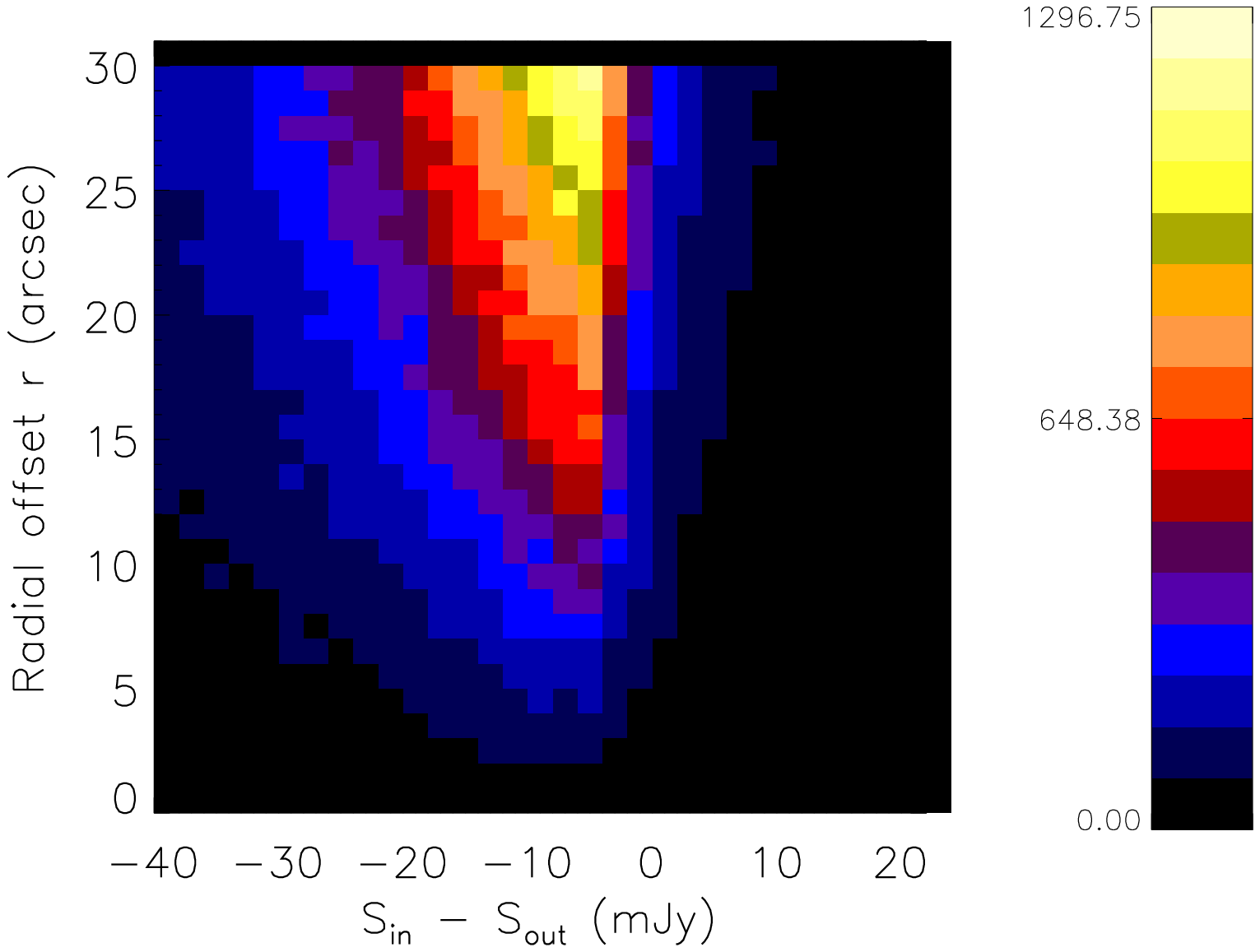}
\includegraphics[width=2.3in]{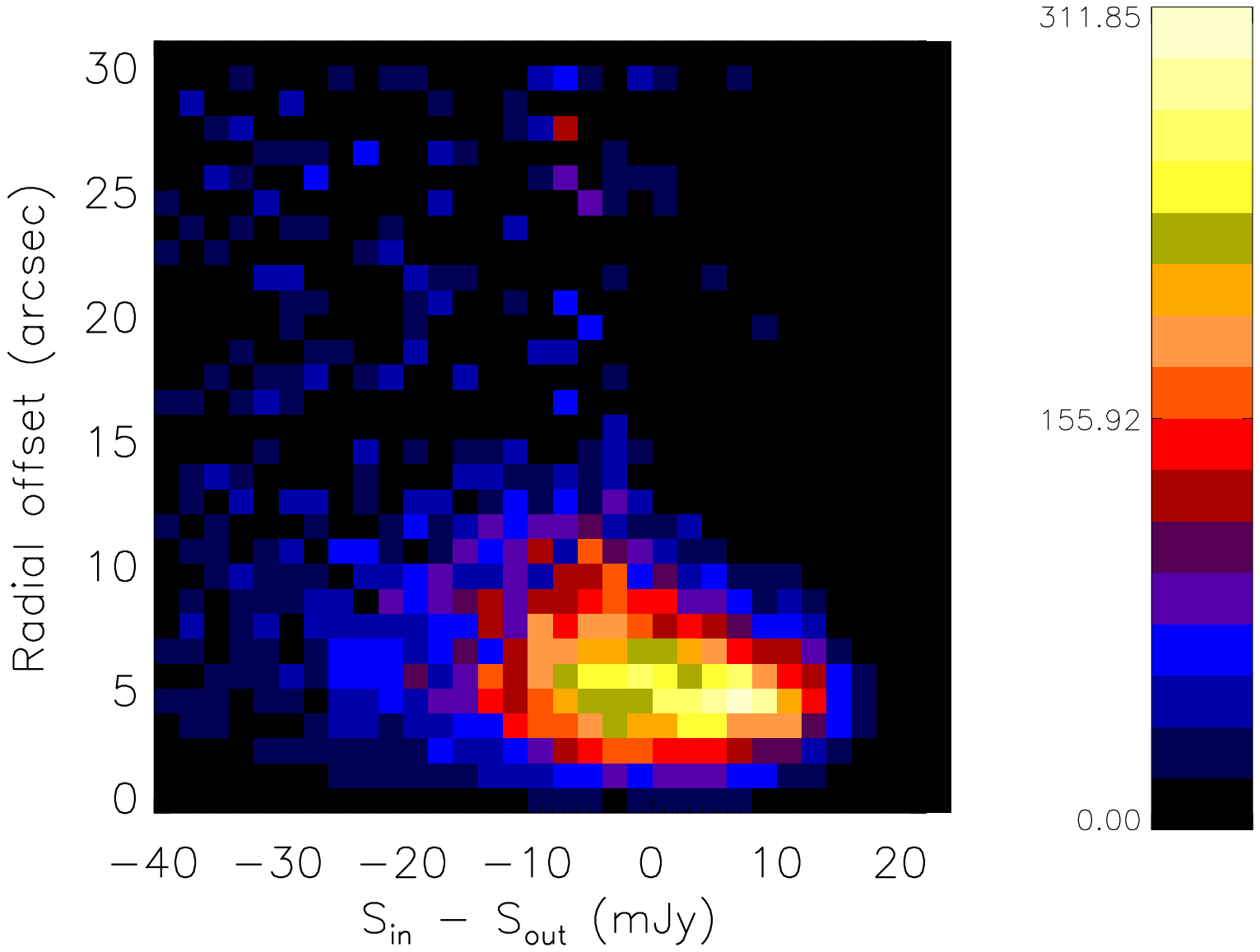}
\includegraphics[width=2.3in]{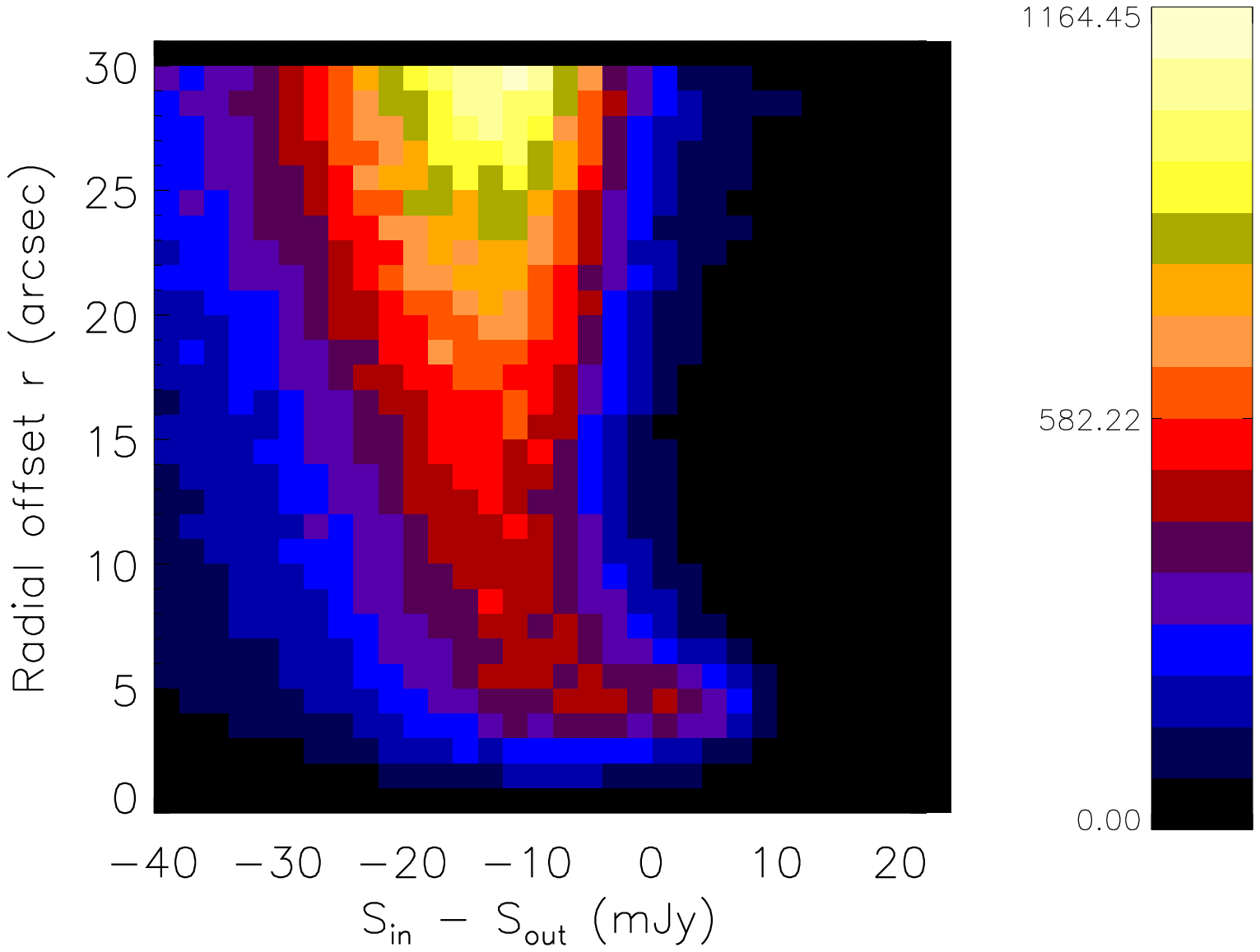}
\includegraphics[width=2.3in]{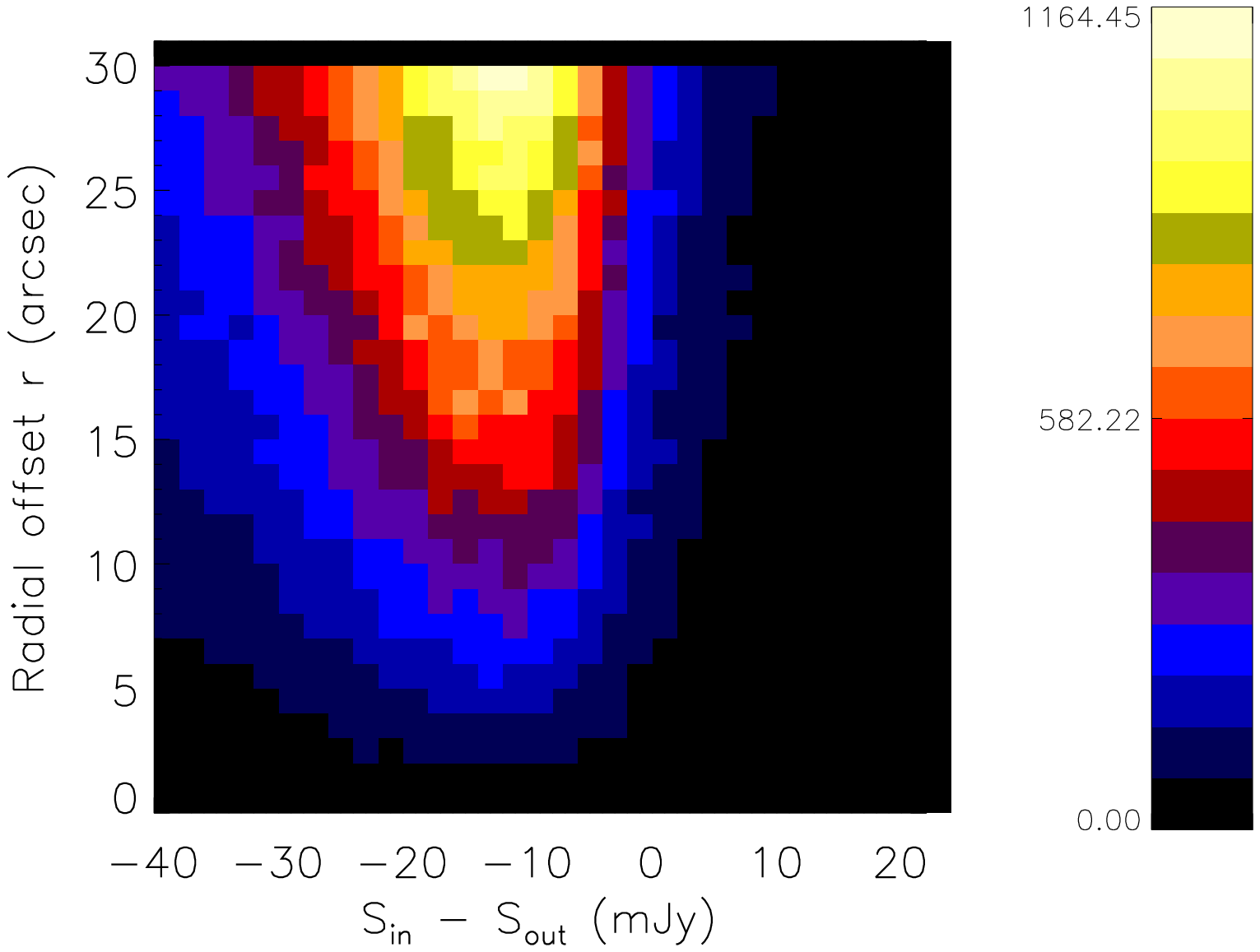}
\includegraphics[width=2.3in]{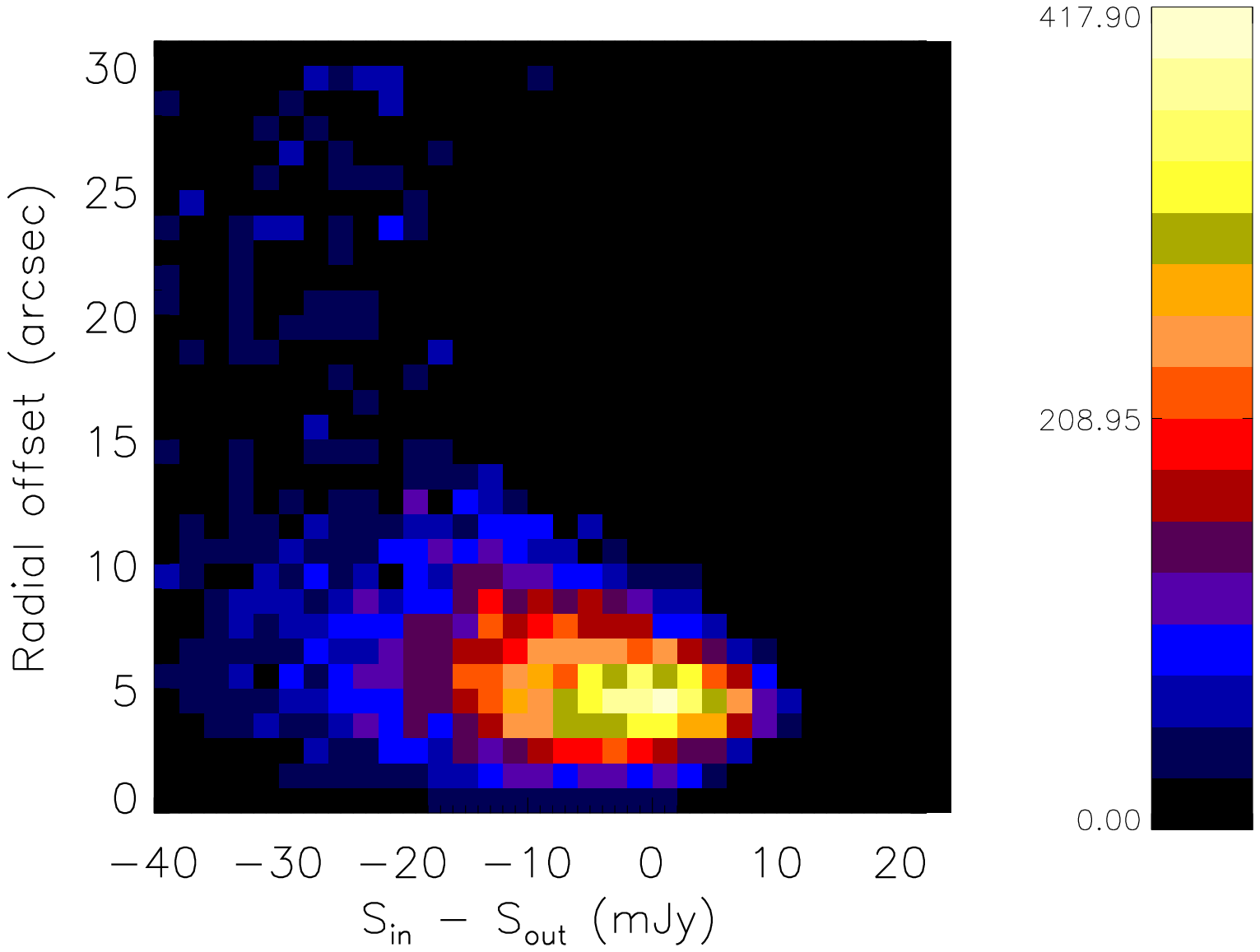}
\caption{The 2D density distribution as a function of radial offset $r$ and flux difference $S_{\rm in} - S_{\rm out}$ in the simulated unclustered COSMOS field at 250 $\micron$ (SXT: top panels; SF: bottom panels). The left panels show the density distribution of all matches between the input and output catalogue. The middle panels show the density distribution of all matches between the input and the randomised output catalogue. The right panels show the difference between the left and middle panels (note the changing colour scale) which can be approximated as the 2D density distribution of the real input-output matches.}
\label{2d_prior}
\end{figure*}

\subsection{End-to-end realistic simulations}

To understand the characteristics of the point source catalogues, we need to make use of realistic simulations of {\it Herschel}-SPIRE observations. We give a brief summary of the steps taken to produce the simulations below:
\begin{enumerate}

\item Generate a SPIRE input source catalogue based on the B{\'e}thermin et al. (2010) source count model with flux densities at 250, 350 and 500 $\micron$.

\item Assign random coordinates (x, y) as well as clustered coordinates to the input catalogue generated from step (i) to make mock SPIRE sky maps. The random mock SPIRE sky is straightforward to generate by simply distributing the input sources randomly in the map. Source positions are the same in simulations at 250, 350 and 500 $\micron$. Clustered coordinates are assigned as follows.  First we generate a single background density map with a power spectrum based on the clustering model fit in Viero et al. (2013) (both the one- and two-halo term, but not the Poisson term).  Next we draw positions weighted by the density map and assign source flux densities to each of the three simulated sky map bands for each position.   The resulting simulated maps have sources correlated in position and colour, with power spectra resembling that of clustered dusty star-forming galaxies.

\item Scan the mock sky at 250, 350 and 500 $\micron$ and make time streams. At the same time, add realistic white and 1/f noise to the simulated time streams.

\item Run the time streams through the SMAP map-making pipeline, and then make final simulated maps which resemble the equivalent SMAP maps in the real observations.  At the same time, it takes the input catalogue and map-specific header file and converts the catalogue coordinates from map pixel coordinates (x, y) to (ra, dec) while excluding those sources located outside the map.

\end{enumerate}

In total, we have simulated five different HerMES fields which are Lockman-SWIRE, COSMOS, UDS, ELAIS-S1 and EGROTH, covering a range of depth (Level 2 to Level 6). In Fig.\ref{pofd}, we plot the normalised pixel flux distribution in the real observation, simulated observations (both random and clustered), and jackknife noise map of the Lockman-SWIRE field at  350 $\micron$ (PMW). The jackknife noise map is made by subtracting two independent maps of the same field using the first and second half of the data. Therefore, the jackknife difference map should remove the sky signal and contain only the instrument noise. The overall shape of the histogram in the simulated maps matches very well to the real histogram. The non-Gaussianity of the real or simulated pixel histograms is due to the presence of point sources as well as the variation of the instrument noise level across the map (which results in a sum of Gaussian distributions). The latter is evident in the pixel histogram of the jackknife map.

In the following sections, we will mostly show results (e.g. positional and photometric accuray) from the simulated COSMOS and Lockman-SWIRE field. The other three fields exhibit similar overall trends.

\subsection{Matching input with output}

In the confusion-limited regime, matching the input catalogue with the output catalogue is far from trivial. On one hand, one detection in the output can result from blending of several input sources. On the other hand, one input source can sometimes contribute to more than one detection in the output.

\begin{figure}\centering
\includegraphics[height=2.5in, width=3.45in]{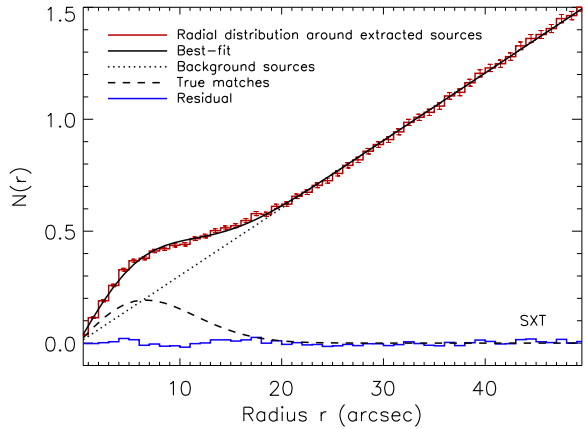}
\caption{The radial distribution of positional offsets between extracted sources and input sources per extracted source (red line), the best-fit model prediction (black solid line), and the difference between the two (blue line). The radial distribution of background input sources uncorrelated with the extracted source grows linearly with $r$. The radial distribution of true matches between the input and output follows a Rayleigh distribution. Note that in this plot we use the SXT 350 $\micron$ catalogue extracted from the unclustered simulation of the COSMOS field.} 
\label{positional_prior}
\end{figure}

\begin{table}
\caption{The $\sigma_r$ value in the Rayleigh radial probability distribution averaged over five different simulated fields at 250, 350 and 500 $\micron$ for SXT and SF catalogues. The top two rows correspond to simulations with randomly distributed input sources, while the bottom two rows correspond to simulations with clustered input sources. In all cases, the SF catalogues have slightly better positional accuracy than the SXT catalogues. By construction, SF250 source catalogues at 250, 350 and 500 $\micron$ have the same $\sigma_r$ value as the SF catalogues at 250 $\micron$.}
\begin{tabular}{lllll}
\hline
Method & PSW & PMW & PLW \\
\hline
SXT (random)& $5.3\arcsec\pm0.2\arcsec$ & $7.9\arcsec\pm0.7\arcsec$ &$12.7\arcsec\pm0.6\arcsec$ \\
SF (random)  & $4.8\arcsec\pm0.2\arcsec$ & $7.2\arcsec\pm0.4\arcsec$ & $11.5\pm0.5\arcsec$ \\
\hline
SXT (clustered) & $5.3\arcsec\pm0.2\arcsec$ & $7.7\arcsec\pm0.7\arcsec$ &$13.3\arcsec\pm0.8\arcsec$ \\
SF  (clustered) & $4.8\arcsec\pm0.1\arcsec$ & $7.0\arcsec\pm0.3\arcsec$ & $11.8\pm0.4\arcsec$ \\
\hline
\end{tabular}
\label{tab:positional_prob}
\end{table}

\begin{table*}
\caption{The best-fit value and scatter for the parameters in the geometric function (Eq. 8) which describes the ratio of output - input flux difference to input flux as a function of input flux density averaged over five different simulated fields at 250, 350 and 500 $\micron$. The top three rows correspond to simulations with randomly distributed input sources, while the bottom three rows correspond to simulations with clustered input sources. The difference between the clustered and unclustered simulations is small.}
\begin{tabular}{llllllllll}
\hline
Method & $a_0^{250}$ & $a_0^{350}$ & $a_0^{500}$ & $a_1^{250}$& $a_1^{350}$ & $a_1^{500}$  & $a_2^{250}$ & $a_2^{350}$& $a_2^{500}$\\
\hline
SXT (random) & $20.5(4.4)$ & $16.9(3.3)$ & $19.7(4.7)$ & $-1.6(0.1)$ & $-1.5(0.1)$ & $-1.4(0.1)$ & $0.0(0.0)$ & $0.0(0.0)$&$0.0(0.0)$\\
SF (random) & $25.6(7.2)$ & $20.5(4.3)$ & $22.3(4.7)$ & $-1.5(0.0)$ & $-1.5(0.0)$ & $-1.4(0.1)$ & $0.0(0.0)$ & $0.0(0.0)$&$0.0(0.0)$\\
SF250 (random) & $25.6(7.2)$ & $17.2(3.1)$ & $8.7(2.2)$ & $-1.5(0.0)$ & $-1.7(0.1)$ & $-2.0(0.2)$ & $0.0(0.0)$ & $0.0(0.0)$&$0.0(0.0)$\\
\hline
SXT (clustered) & $17.9(9.1)$ & $21.0(4.5)$ & $21.4(5.4)$ & $-1.4(0.4)$ & $-1.7(0.4)$ & $-1.4(0.1)$ & $0.0(0.0)$ & $0.0(0.0)$&$0.0(0.0)$\\
SF (clustered) & $26.9(7.1)$ & $22.3(5.5)$ & $22.9(6.3)$ & $-1.6(0.1)$ & $-1.5(0.1)$ & $-1.4(0.1)$ & $0.0(0.0)$ & $0.0(0.0)$&$0.0(0.0)$\\
SF250  (clustered)  & $26.9(7.1)$ & $18.7(3.6)$ & $9.3(2.7)$ & $-1.6(0.1)$ & $-1.7(0.1)$ & $-1.9(0.1)$ & $0.0(0.0)$ & $0.0(0.0)$&$0.0(0.0)$\\
\hline
\end{tabular}
\label{tab:flux_prob}
\end{table*}

We match the input truth list with the output source list using a likelihood ratio (LR) method similar to Chapin et al. (2011). For each input-output source pair, we calculate the LR, the ratio of probability of being a true match to probability of being a random association,  based on the positional offset $r$ and flux difference $\Delta S = S_{\rm in} - S_{\rm out}$,
\begin{equation}
{\rm LR (\Delta S, r)} = \frac{q(\Delta S) f( r)}{2\pi r \rho(\Delta S)},
\end{equation}
where $f( r )$ is the probability distribution function (PDF) of the true matches between the input and output as a function of positional offset, $2\pi r$ is the positional distribution of the random matches (assuming a constant surface density of random matches), $q(\Delta S)$ is the PDF of the true matches as a function of flux difference, and $\rho(\Delta S)$ is the PDF of the random matches as a function of flux difference.

In Eq. (6), we have assumed that the LR is separable in positional offset and flux difference. In other words, the flux difference distribution has no dependence on the positional offset and vice versa. To check whether this assumption is valid, we can look at the two-dimensional (2D) density distribution of real input-output matches in the radial offset $r$ vs flux difference $\Delta S = S_{\rm in} - S_{\rm out}$ plane. In the left panels in Fig.~\ref{2d_prior}, we plot the 2D density distribution of all matches between the input and output catalogue in the unclustered COSMOS simulation at 250 $\micron$, which include both the real and random matches between the input and output. In the middle panels in Fig.~\ref{2d_prior}, we plot the density distribution of all matches between the input and randomised output catalogue\footnote{The randomised output catalogue is generated by randomly disturbing the source positions and swapping flux densities between sources in the output list.}, which should only include random matches. The difference between the left and the middle panels, plotted in the right panels in Fig.~\ref{2d_prior},  can be approximated as the density distribution of the real input-output matches. We can see that the flux difference distribution does not depend on the radial offset significantly and the radial offset distribution does not change significantly with the flux difference either. Therefore, the separation of positional offset and flux difference in the LR calculation in Eq. (6) is justified. Simulations with randomly distributed input sources are used in Fig.~\ref{2d_prior}, but the same conclusion that the distribution of positional offset and flux difference can be separated also holds for simulations with clustered input sources.

For the positional PDF of the real input-output matches, we assume a symmetric Gaussian distribution as a function of orthogonal positional coordinates. So, $f( r )$ follows a Rayleigh radial probability distribution, 
\begin{equation}
f(r) = \frac{r}{\sigma^2_r} \exp(-r^2/2\sigma^2_r).
\end{equation}
The positional distribution of random sources uncorrelated with the output sources follows a linear trend with the radial offset $r$ assuming a constant surface density of background sources. In Fig.~\ref{positional_prior}, we plot the histogram of radial offsets for all possible pairs between the input and output per output source within 50\arcsec. The histogram can be fit by the sum of the true matches following a Rayleigh distribution and the random sources following a linear trend with $r$. Poisson errors in the histogram are used in the fitting procedure. We checked that bootstrap errors are very similar to the Poisson errors and do not change the fit. In Fig.~\ref{positional_prior}, we use the SXT 350 $\micron$ catalogue extracted from the unclustered simulation of the COSMOS field. Similar trends are found in other simulations at other wavelengths. In Table 3, we list the best-fit values and uncertainties for $\sigma_r$ for SXT and SF catalogues averaged over all five simulated fields at 250, 350 and 500 $\micron$ respectively. The difference between the clustered and unclustered simulations is very small at all wavelengths. By construction, SF250 source catalogues at 250, 350 and 500 $\micron$ have the same $\sigma_r$ value as the SF catalogues at 250 $\micron$. The SF catalogues have a smaller $\sigma_r$ than the SXT catalogues at all wavelengths, which is expected as SF optimises the source positions during the local fitting process.

\begin{figure*}
\centering
\includegraphics[width=2.3in]{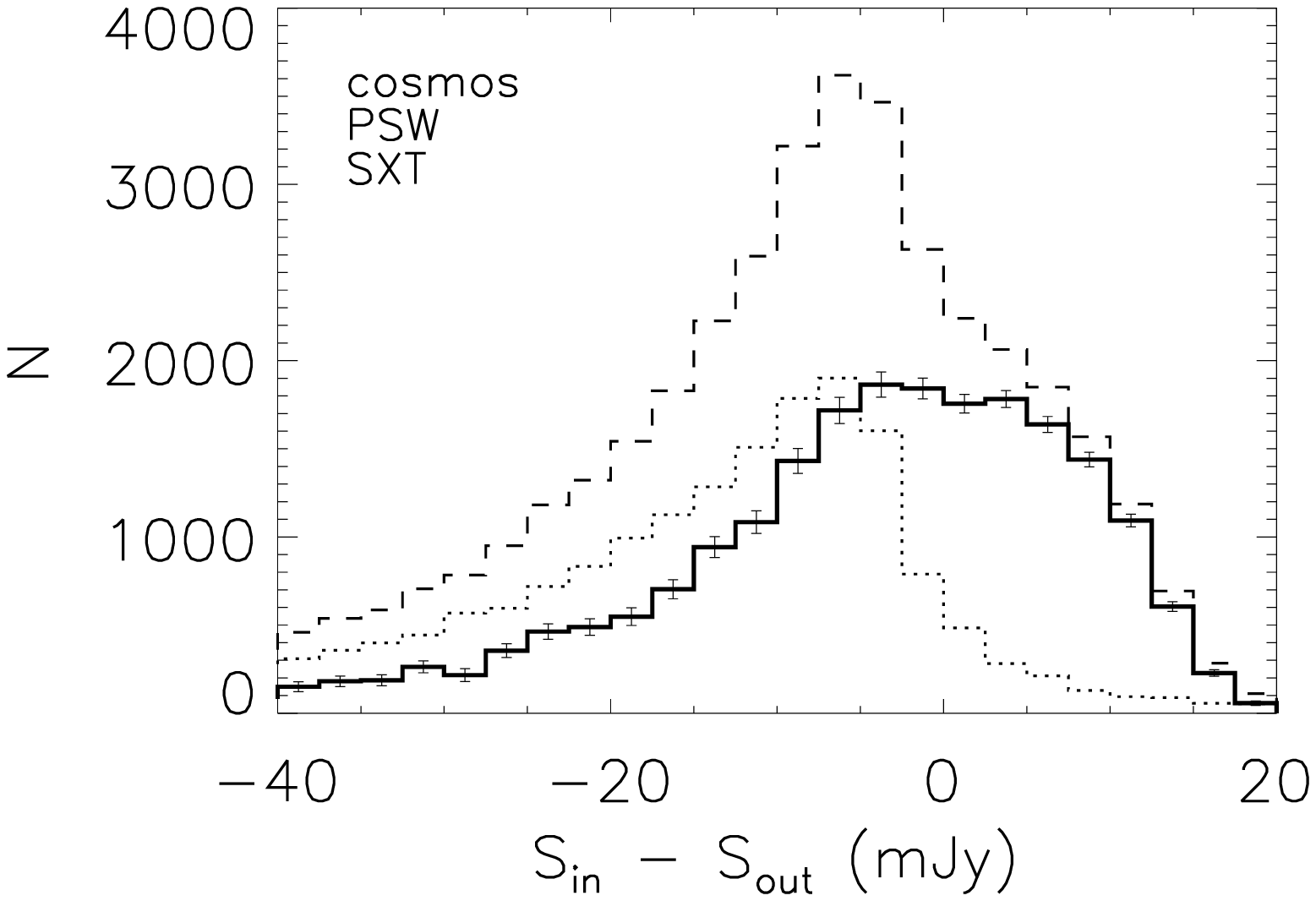}
\includegraphics[width=2.3in]{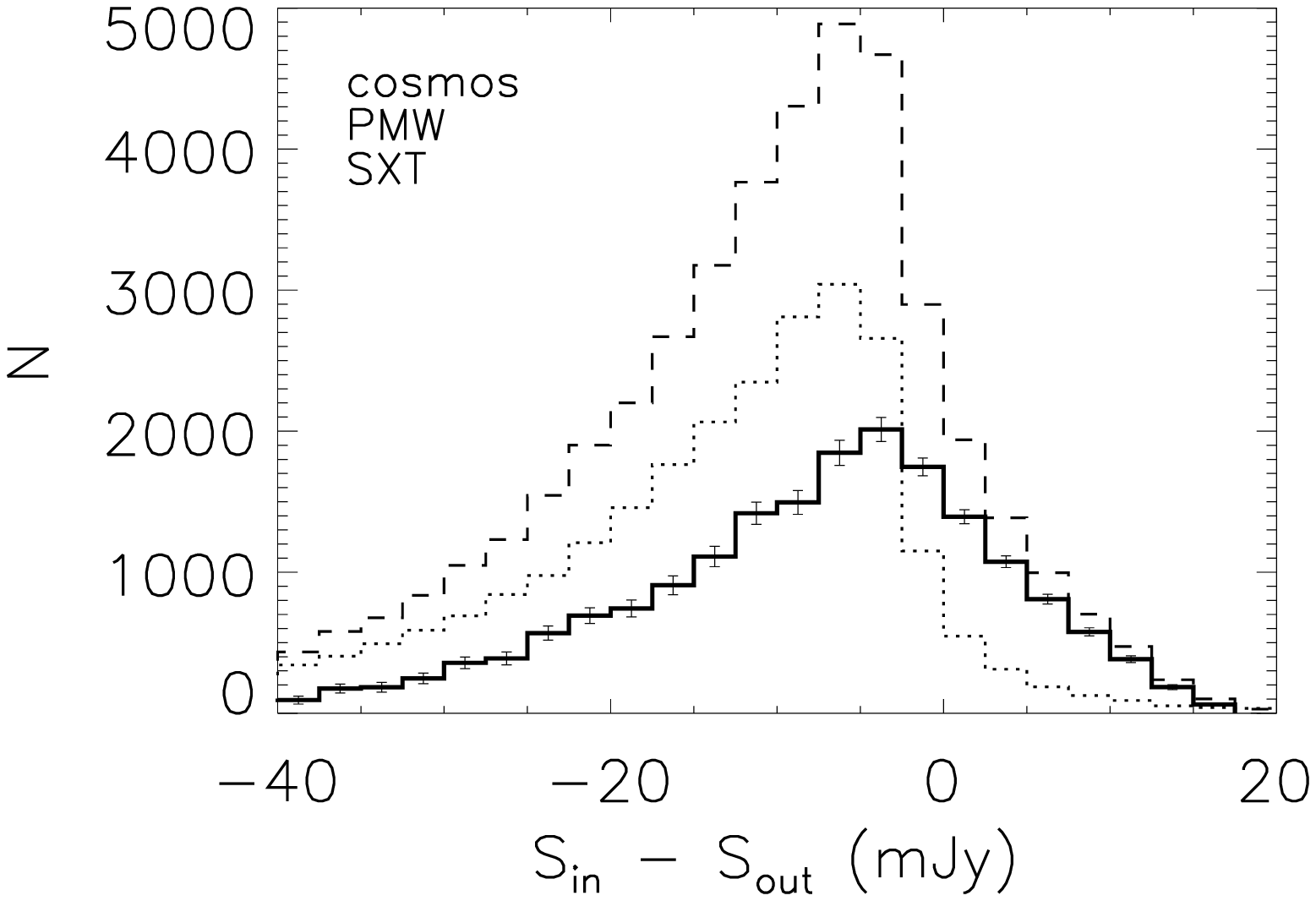}
\includegraphics[width=2.3in]{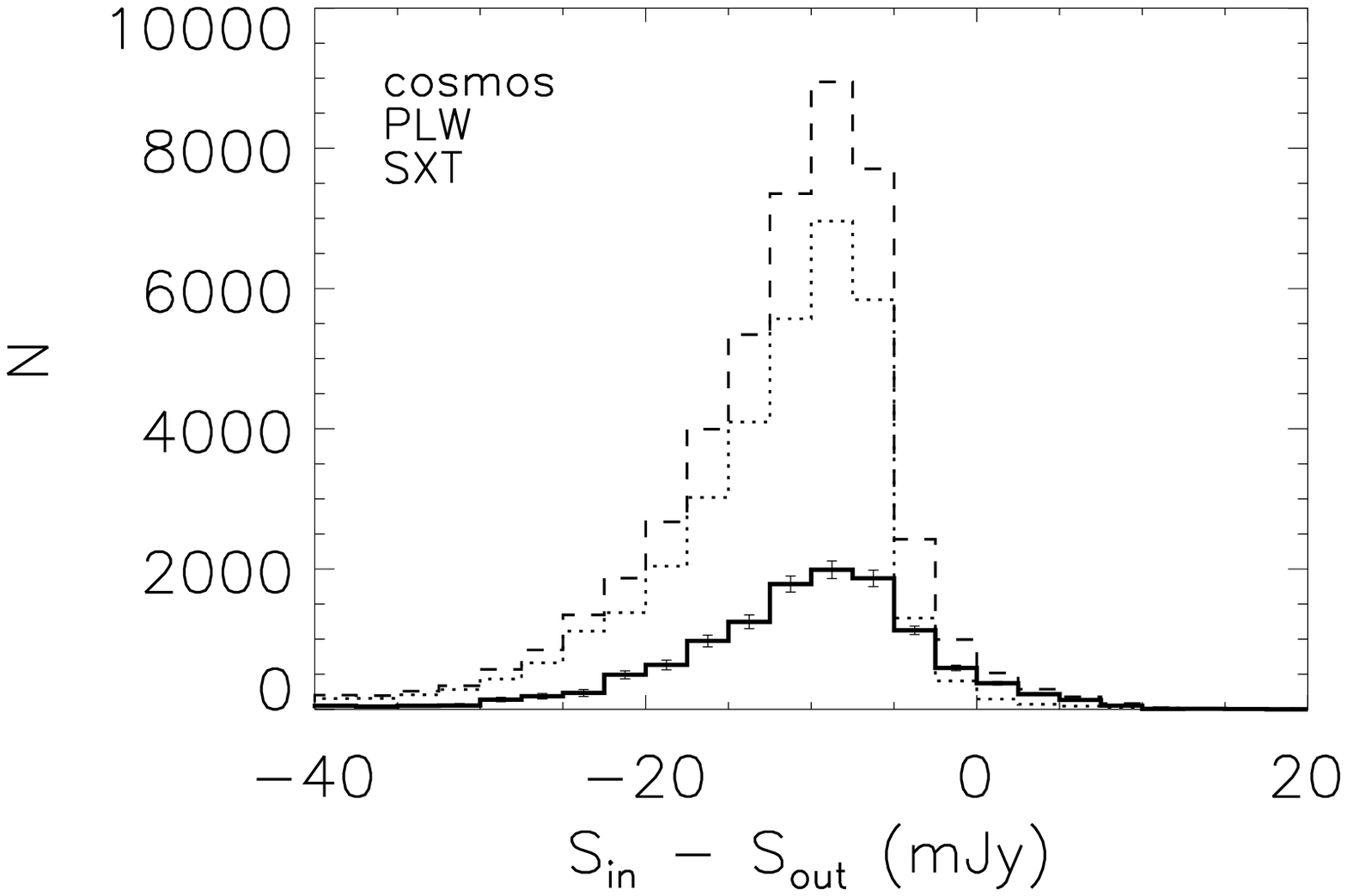}
\includegraphics[width=2.3in]{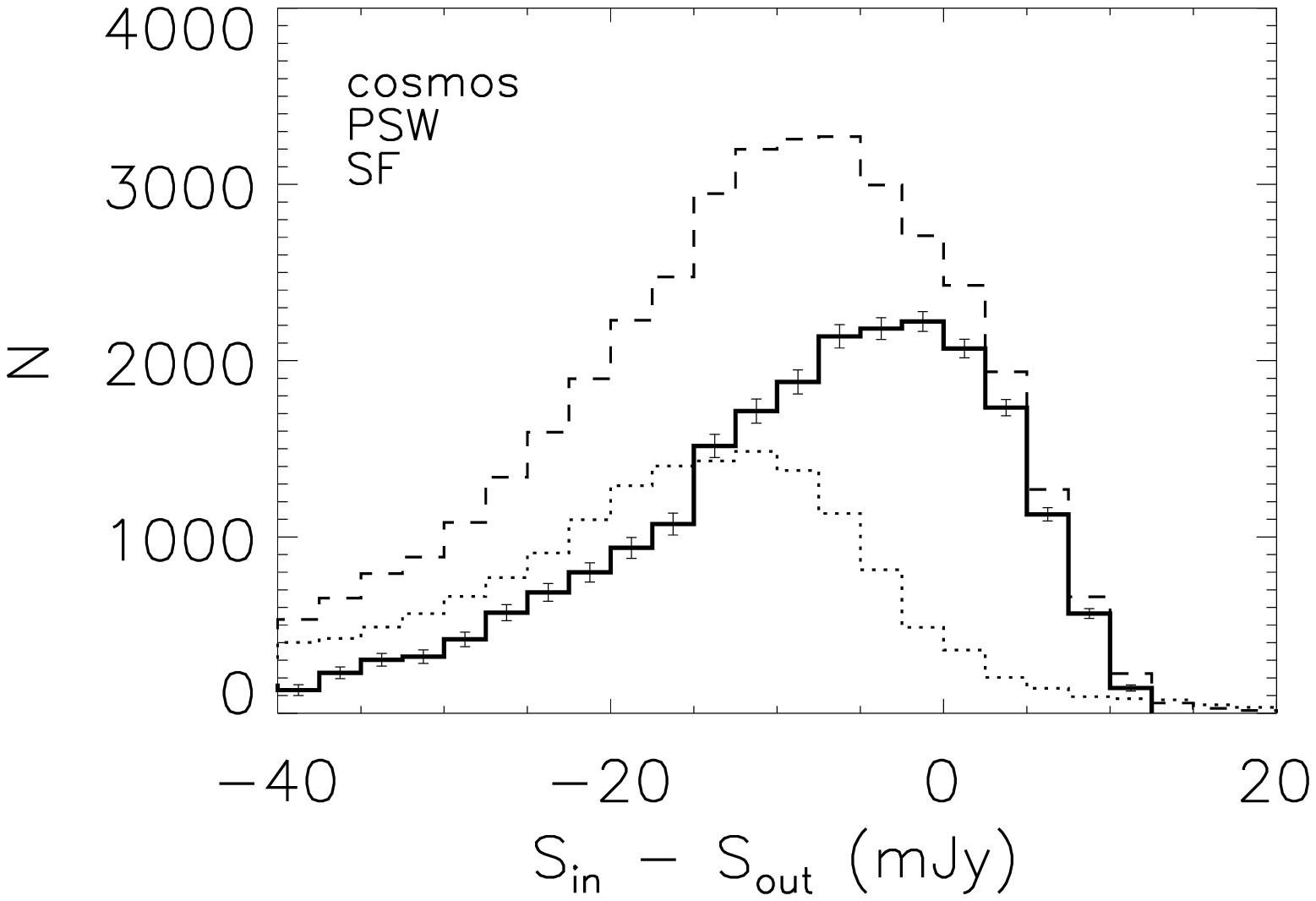}
\includegraphics[width=2.3in]{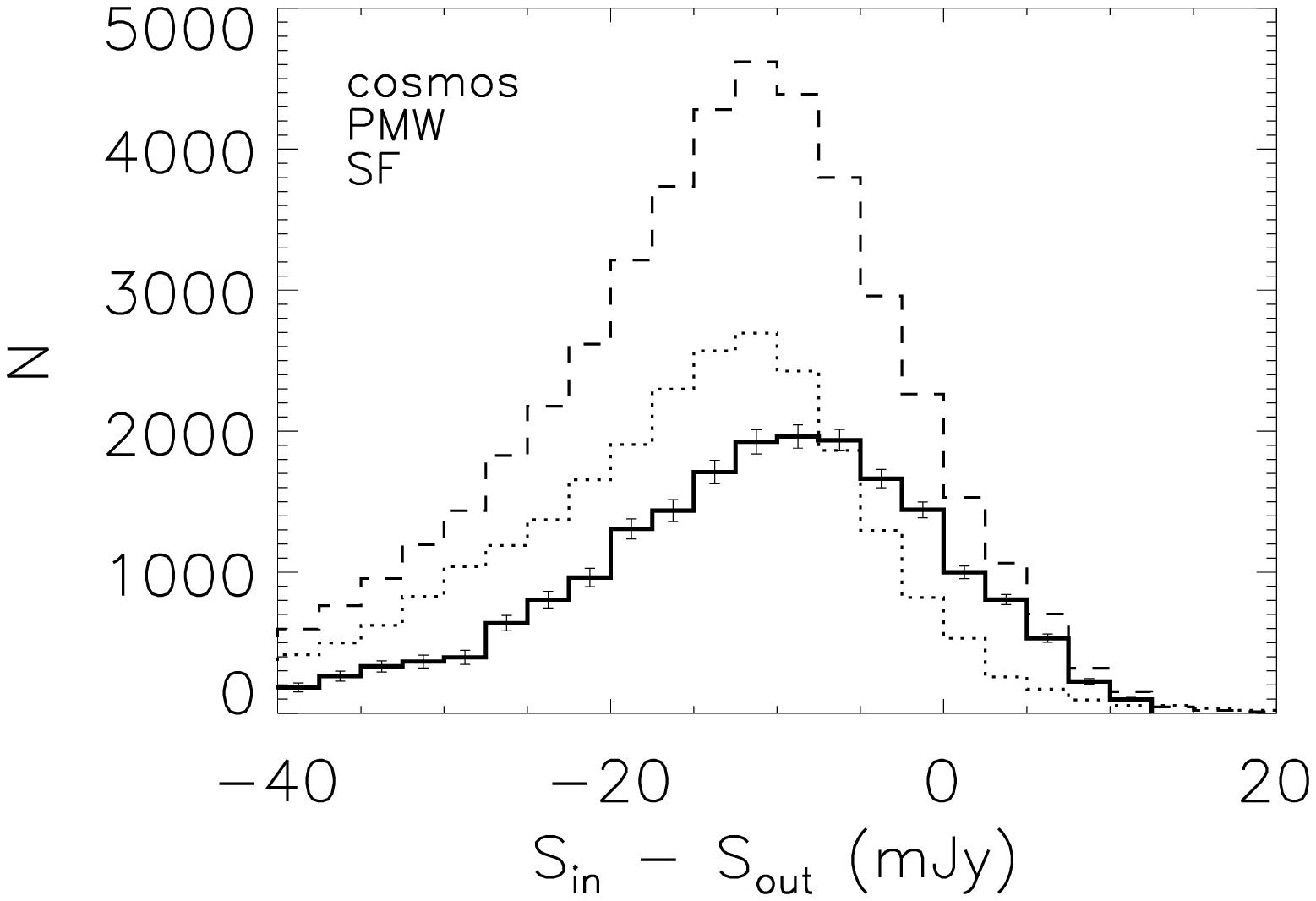}
\includegraphics[width=2.3in]{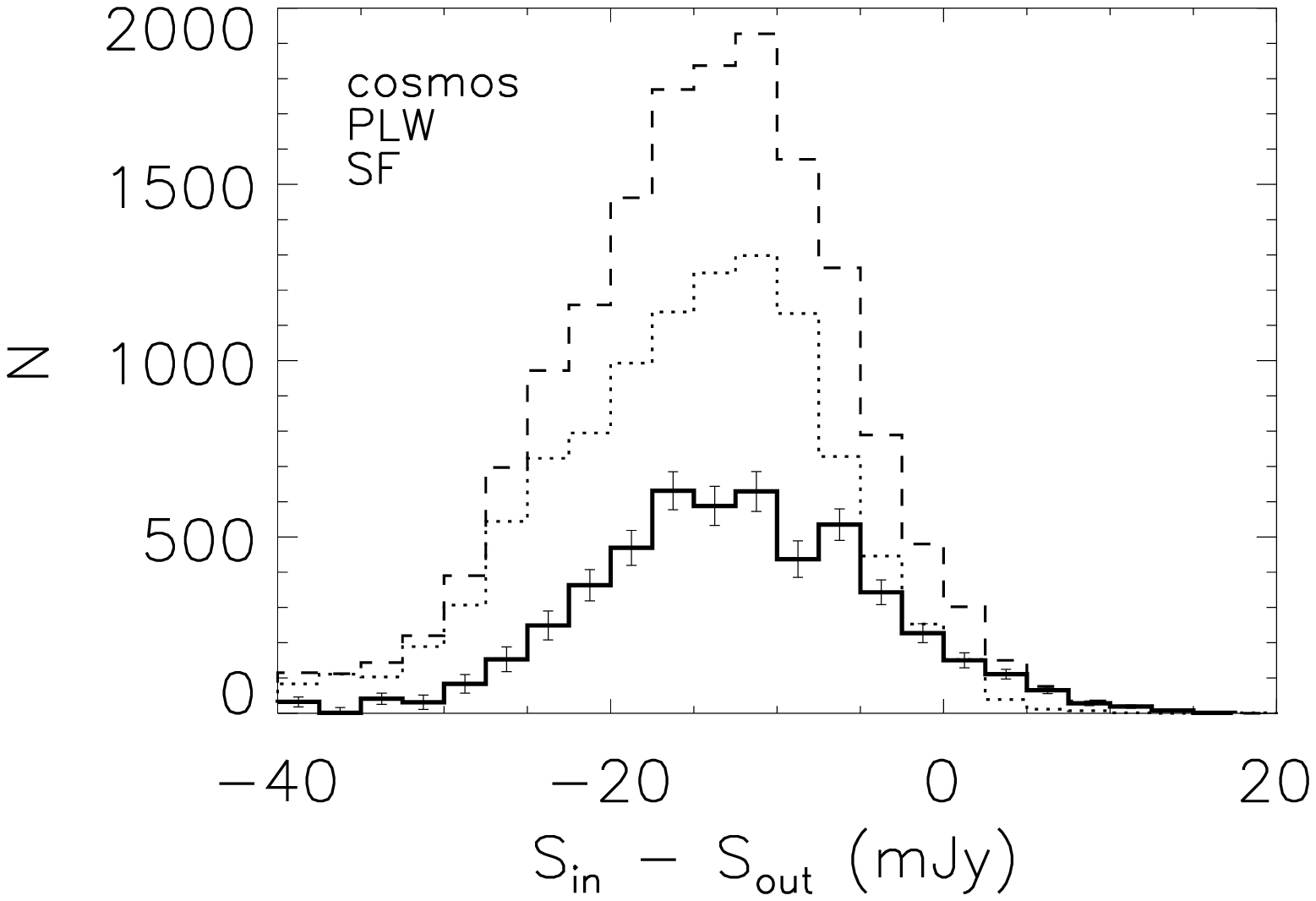}
\includegraphics[width=2.3in]{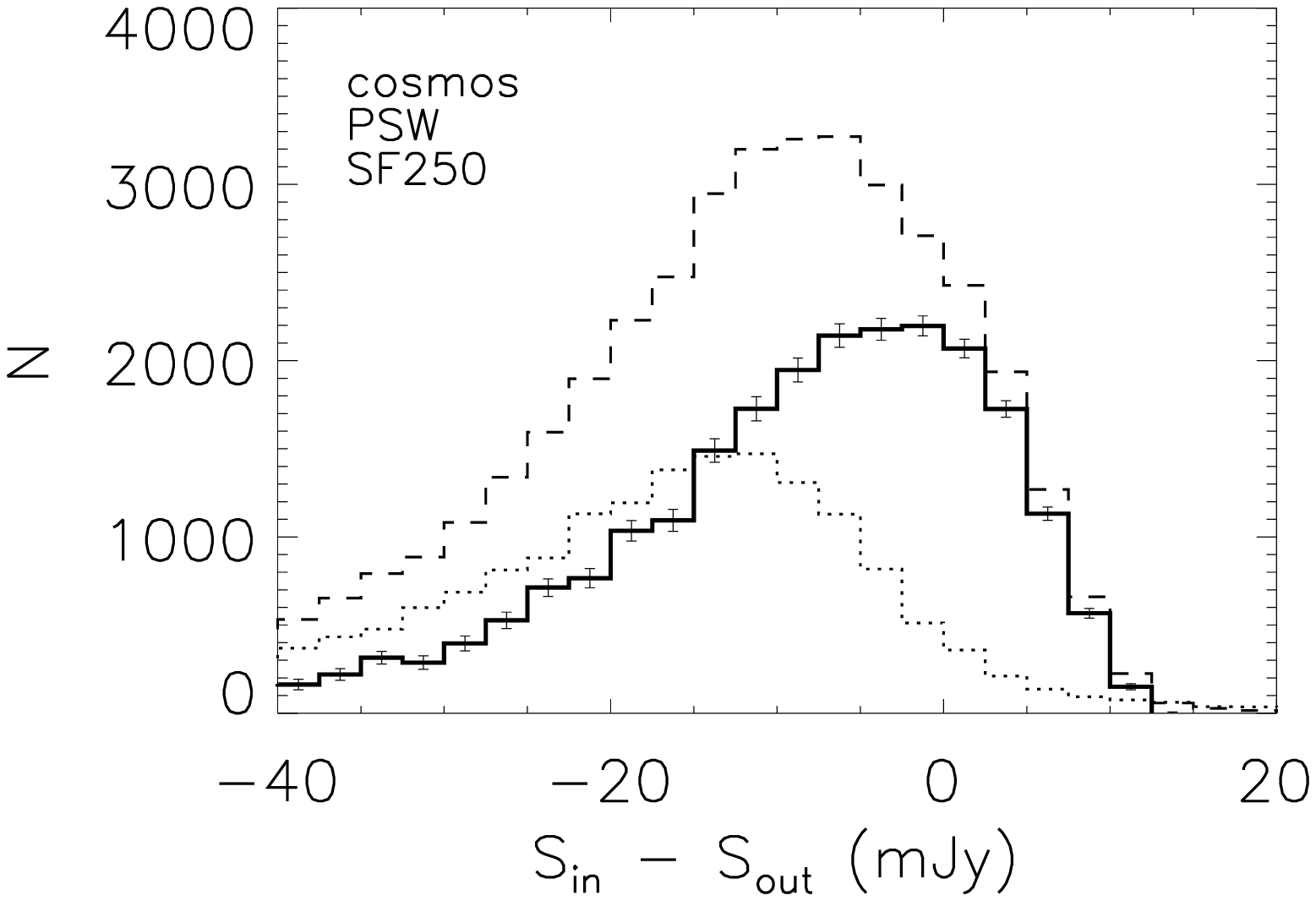}
\includegraphics[width=2.3in]{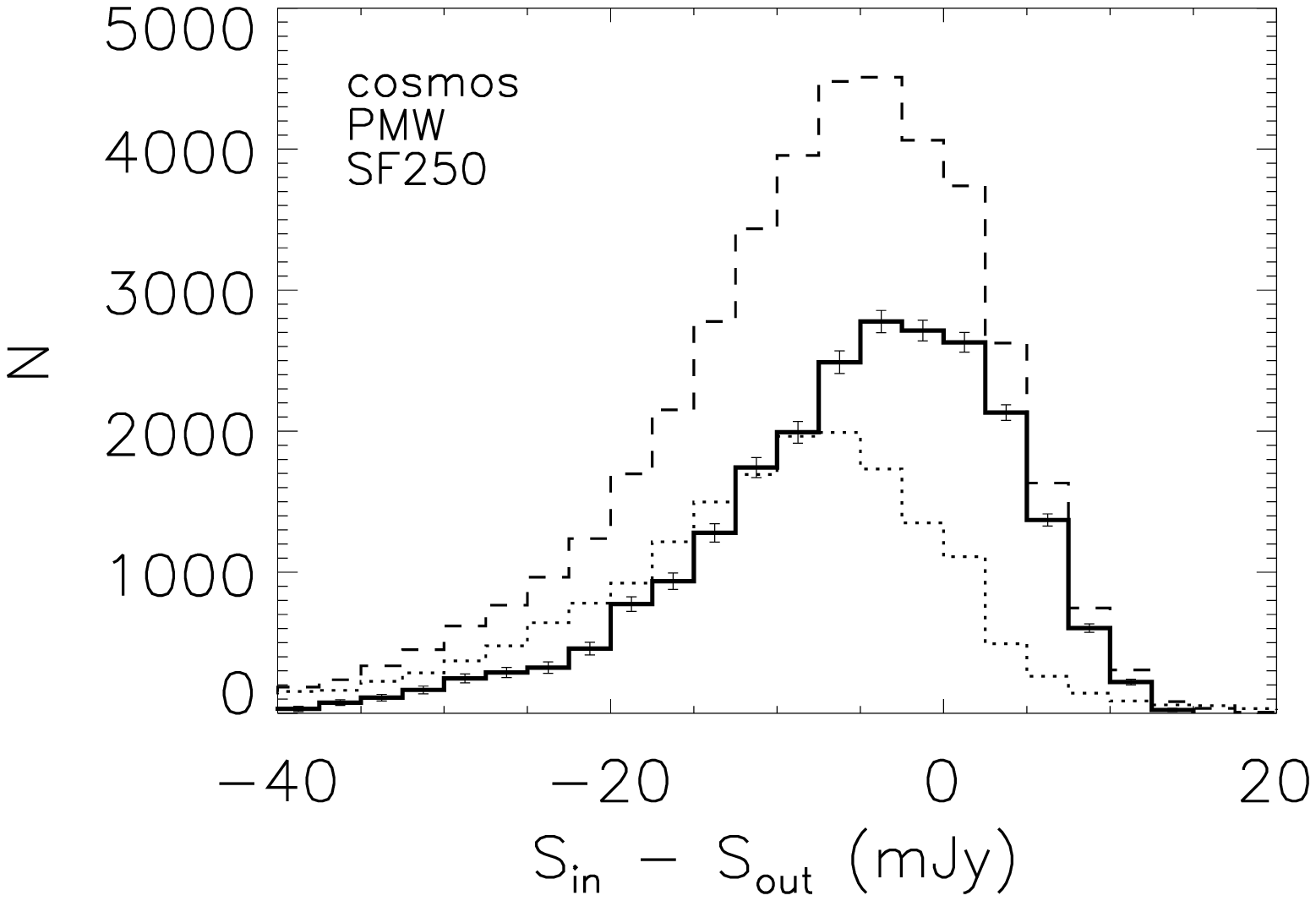}
\includegraphics[width=2.3in]{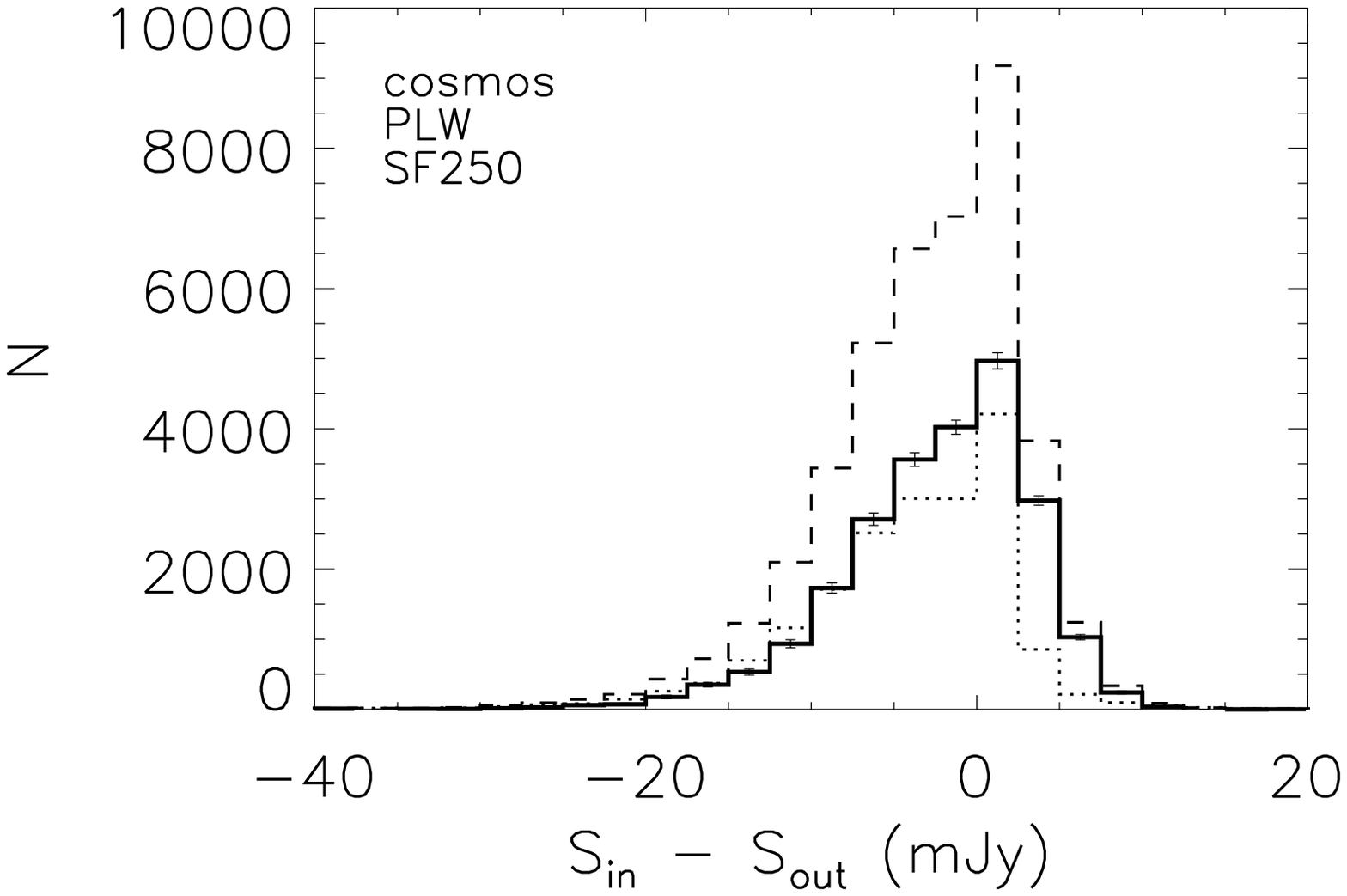}
\caption{The flux difference distribution of all matches between the input and output (dashed histogram) in the simulated unclustered COSMOS field, between the input and randomised output (dotted histogram), and the difference between the two (solid histogram) for SXT (top panels), SF (middle panels) and SF250 (bottom panels) catalogues. Each column corresponds to a different band (left: 250 $\micron$ (PSW); middle: 350 $\micron$ (PSW); right: 500 $\micron$ (PLW)).}
\label{fluxdiff_prior}
\end{figure*}

Next, we need to determine the PDF of the true matches and random matches between the input and output as a function of flux difference, i.e. $q(\Delta S)$ and $\rho(\Delta S)$. It is straightforward to determine $\rho(\Delta S)$. We simply match the input list with the randomised output catalogue and derive the number of random matches as a function of $\Delta S$. To determine $q(\Delta S)$,  first we need to identify the search radius within which the signal-to-noise of the true matches is highest. Using the optimal search radii, we can then derive the histogram of the flux difference for all matches between the input and output. In Fig.~\ref{fluxdiff_prior}, we plot the flux difference distribution of all matches between the input and output within the optical search radii, the flux difference distribution for all matches between the input and randomised output, and the difference between the two (i.e. $q(\Delta S)$) for the SXT, SF and SF250 source catalogues in the simulated unclustered COSMOS field. Errors on the flux difference distribution correspond to Poisson noise. We can see that for both SXT and SF catalogues, the peak of $q(\Delta S)$ shifts to lower values of $\Delta S$ from PSW to PLW, as a result of more severe blending as the beam size increases. The flux difference distribution of the real matches for the SF250 catalogues peaks much closer to zero compared to the SXT and SF catalogues at 350 and 500 $\micron$. This is because the input SF catalogue extracted from the 250 $\micron$ map significantly reduces the level of confusion noise  at 350 and 500 $\micron$.

\begin{figure*}
\centering
\includegraphics[height=2.3in, width=3.45in]{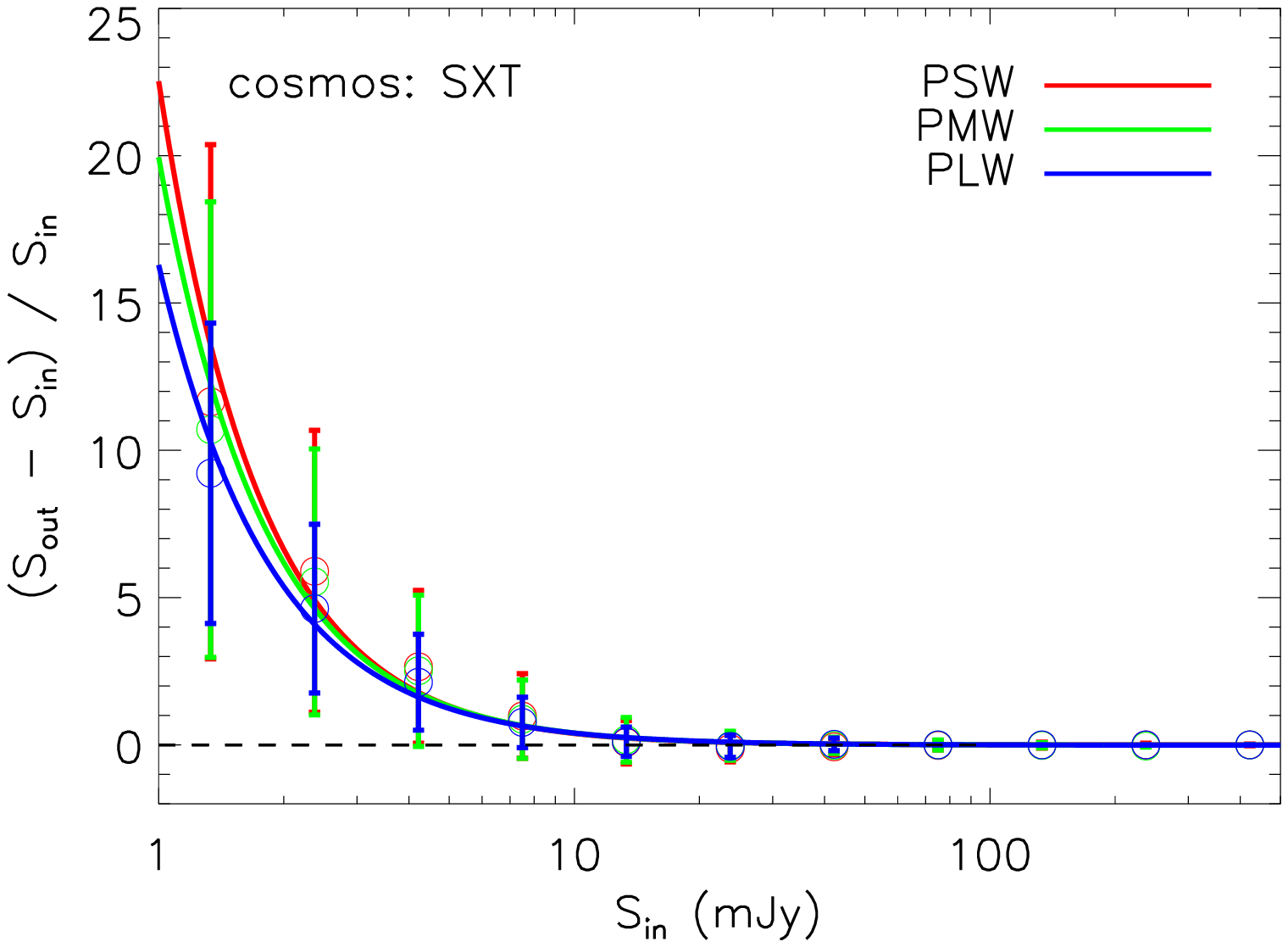}
\includegraphics[height=2.3in, width=3.45in]{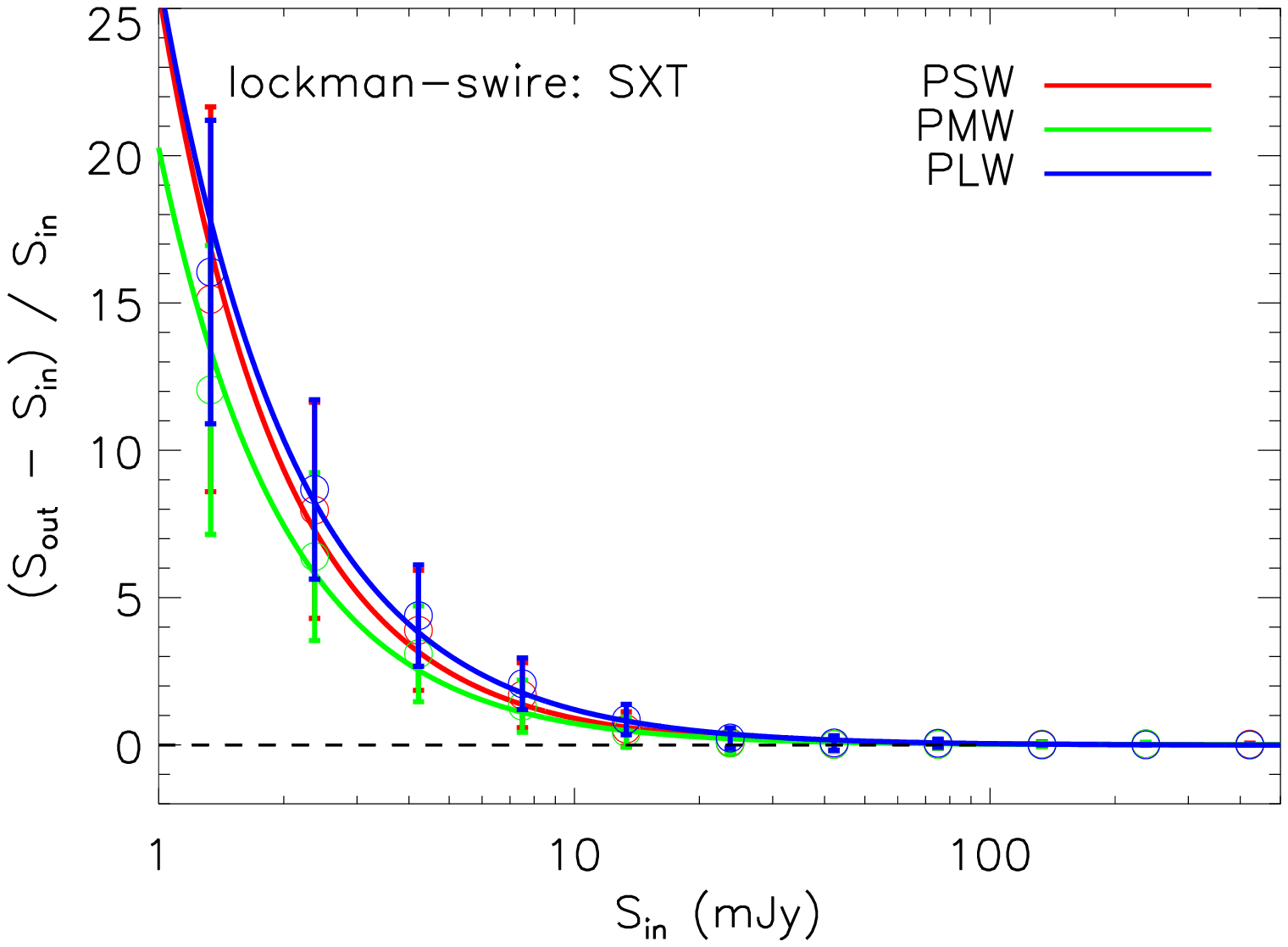}
\includegraphics[height=2.3in, width=3.45in]{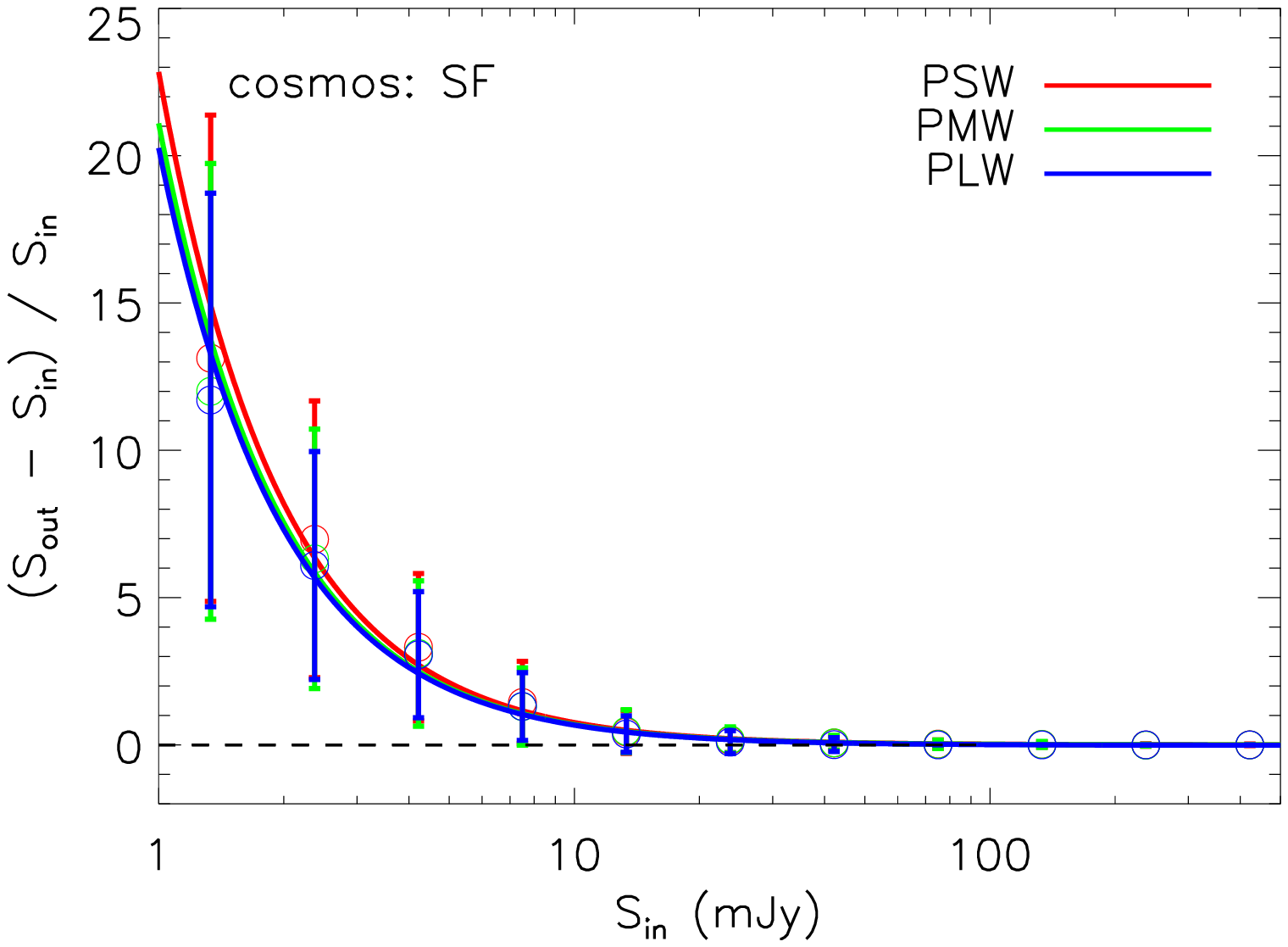}
\includegraphics[height=2.3in, width=3.45in]{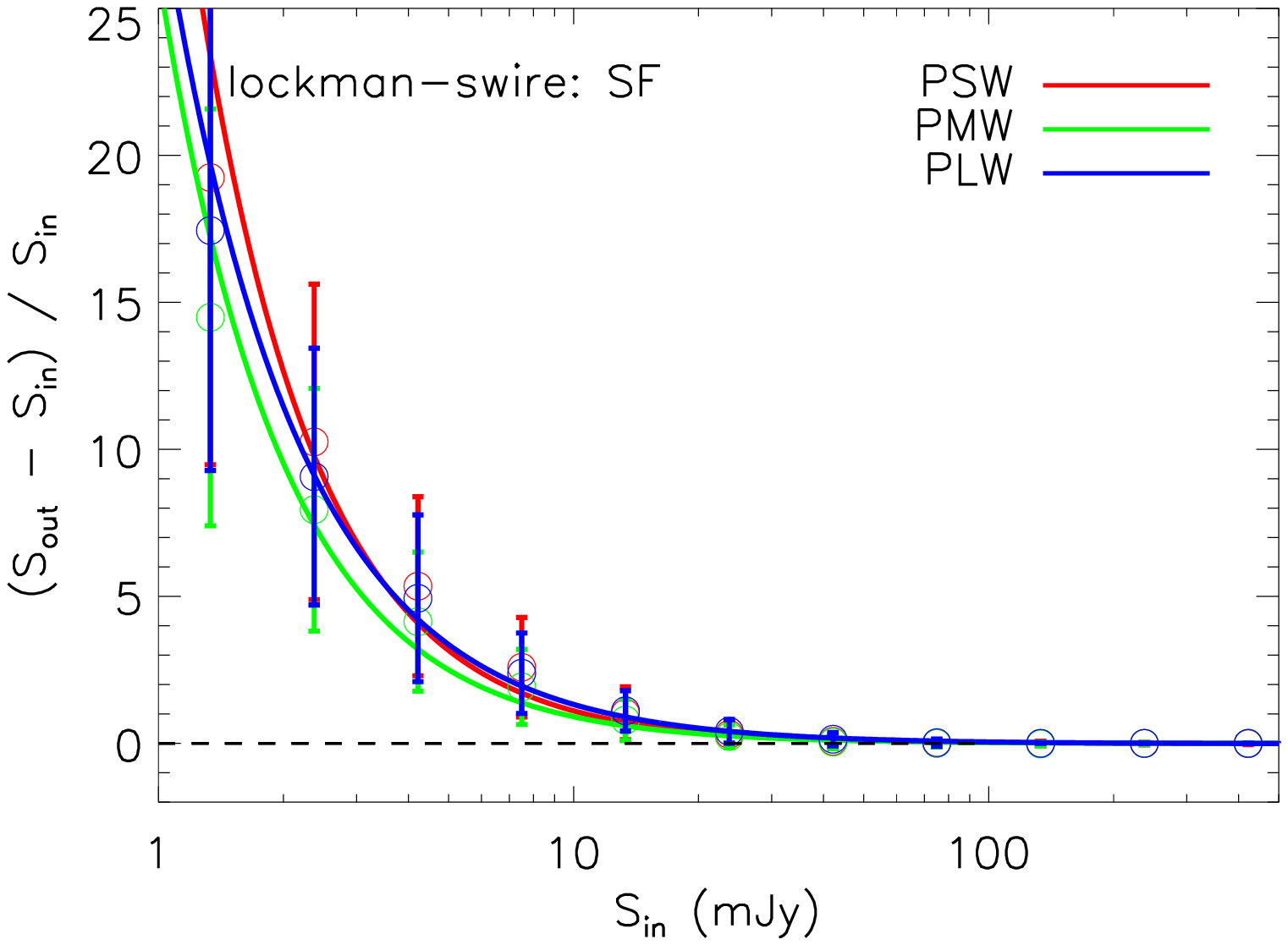}
\includegraphics[height=2.3in, width=3.45in]{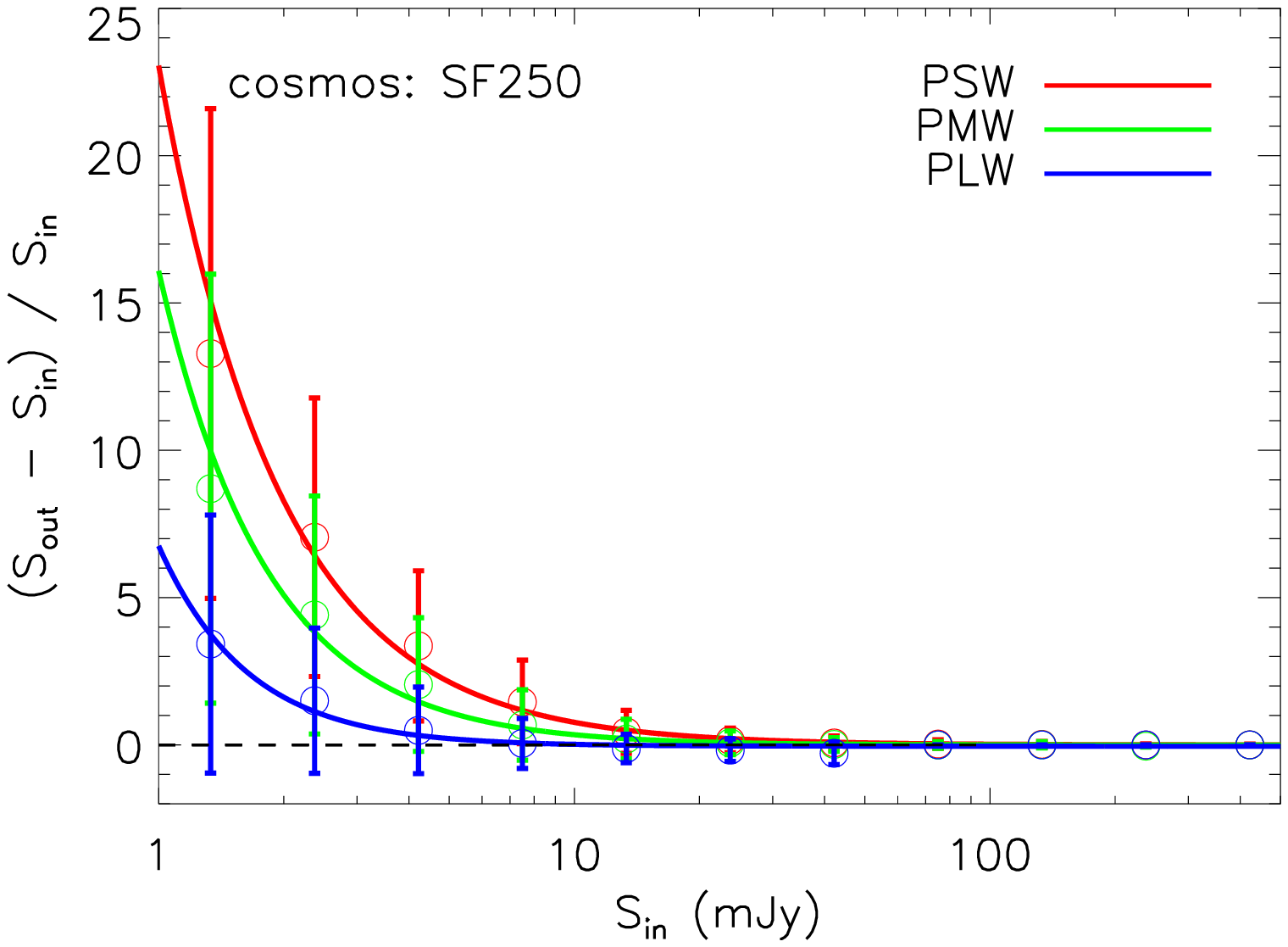}
\includegraphics[height=2.3in, width=3.45in]{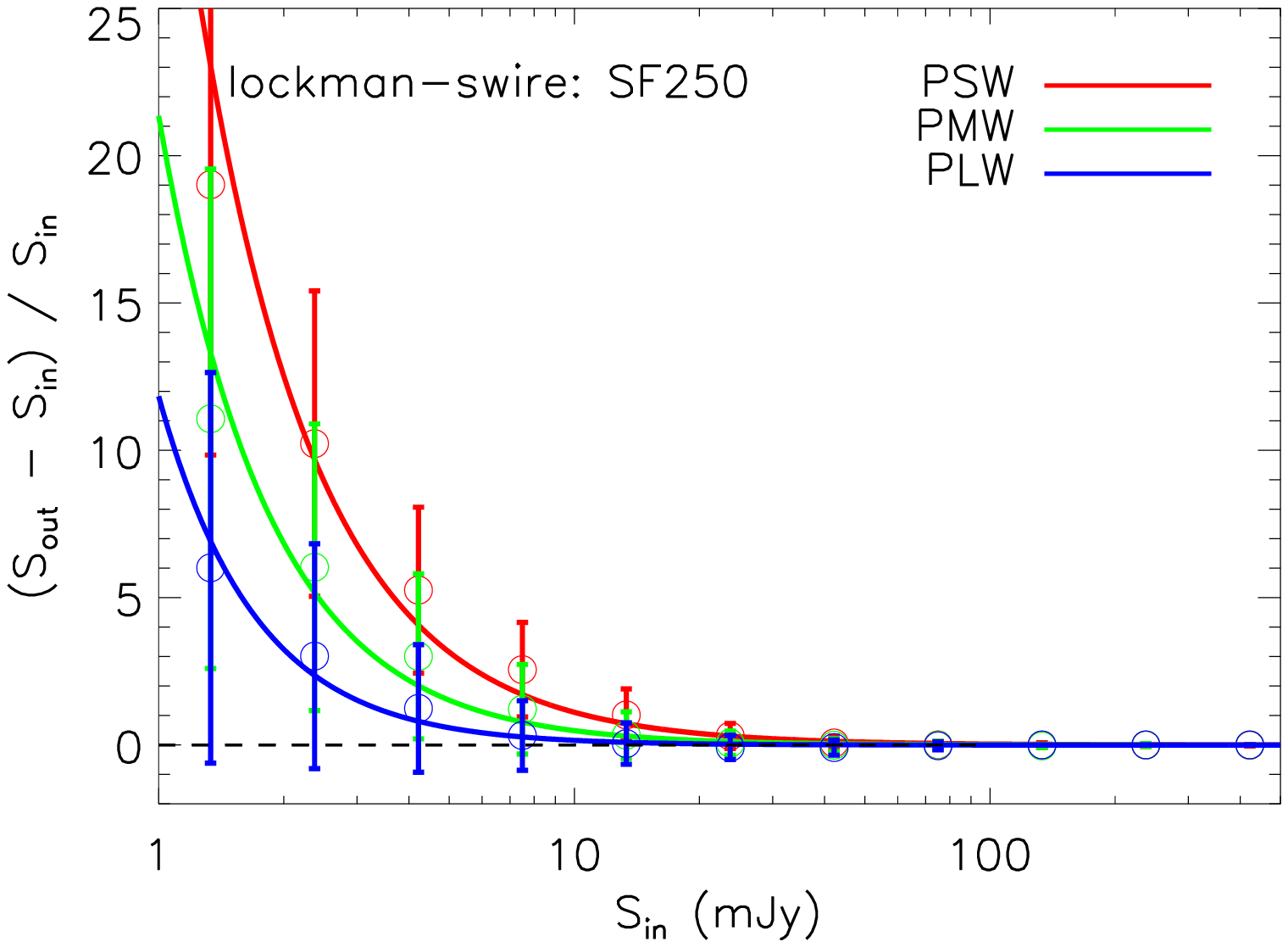}
\caption{The flux difference (output flux - input flux) to input flux ratio as a function of input flux density at 250, 350 and 500 $\micron$ for the three different types of source catalogues (top: SXT; middle: SF; bottom: SF250). The horizontal dashed line corresponds to $S_{\rm out} = S_{\rm in}$. The left column corresponds to the simulated unclustered COSMOS field and the right column corresponds to the simulated unclustered Lockman-SWIRE field. The curves are the best-fit geometric functions (Eq. 8) which describe the ratio of output - input flux difference as a function of input flux density. The best-fit coefficients averaged over all five simulated fields are listed in Table 4. Simulations with clustered input sources produce similar curves.}
\label{fluxunc}
\end{figure*}

\begin{figure*}
\centering
\includegraphics[height=2.3in, width=3.3in]{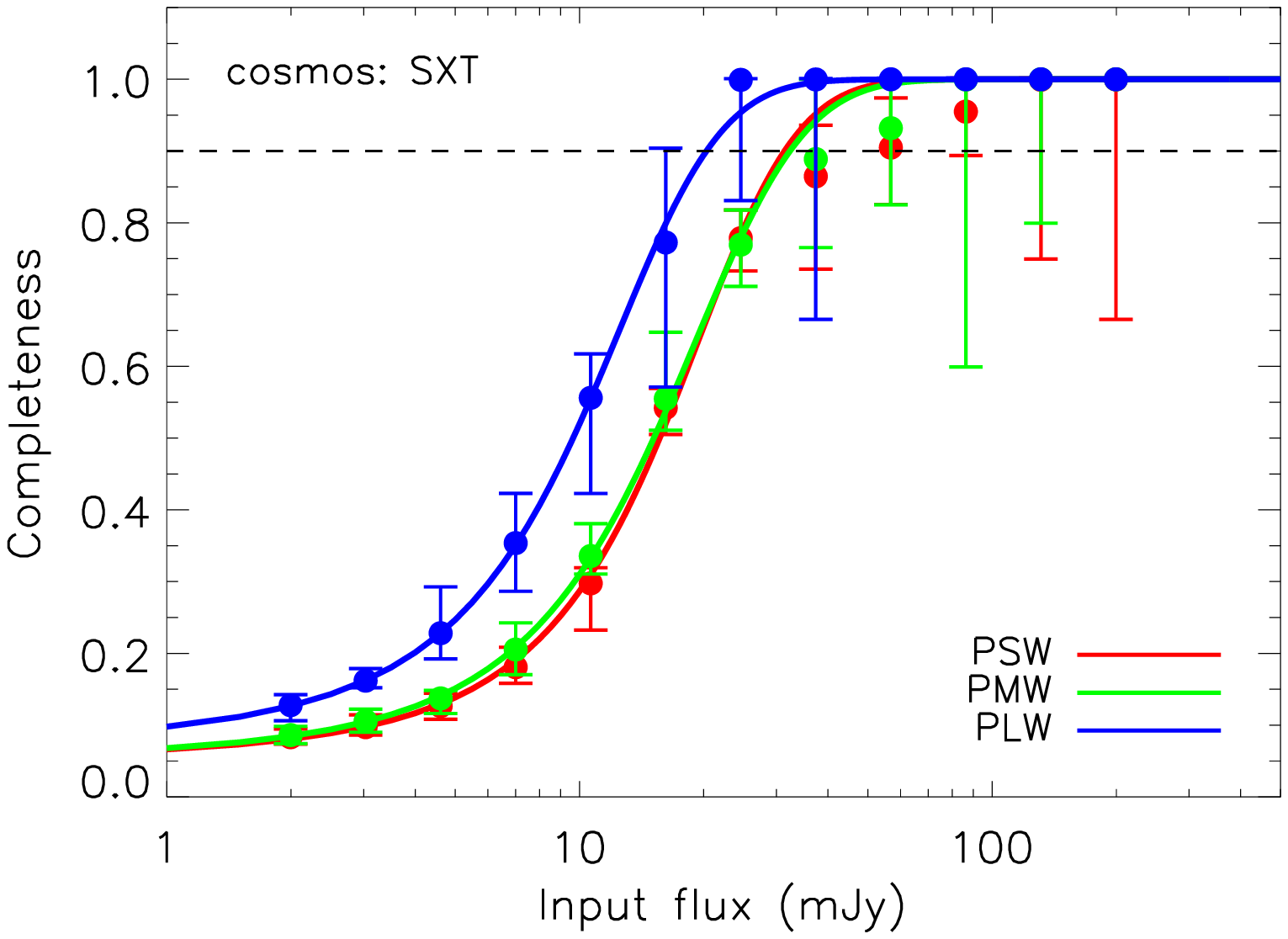}
\includegraphics[height=2.3in, width=3.3in]{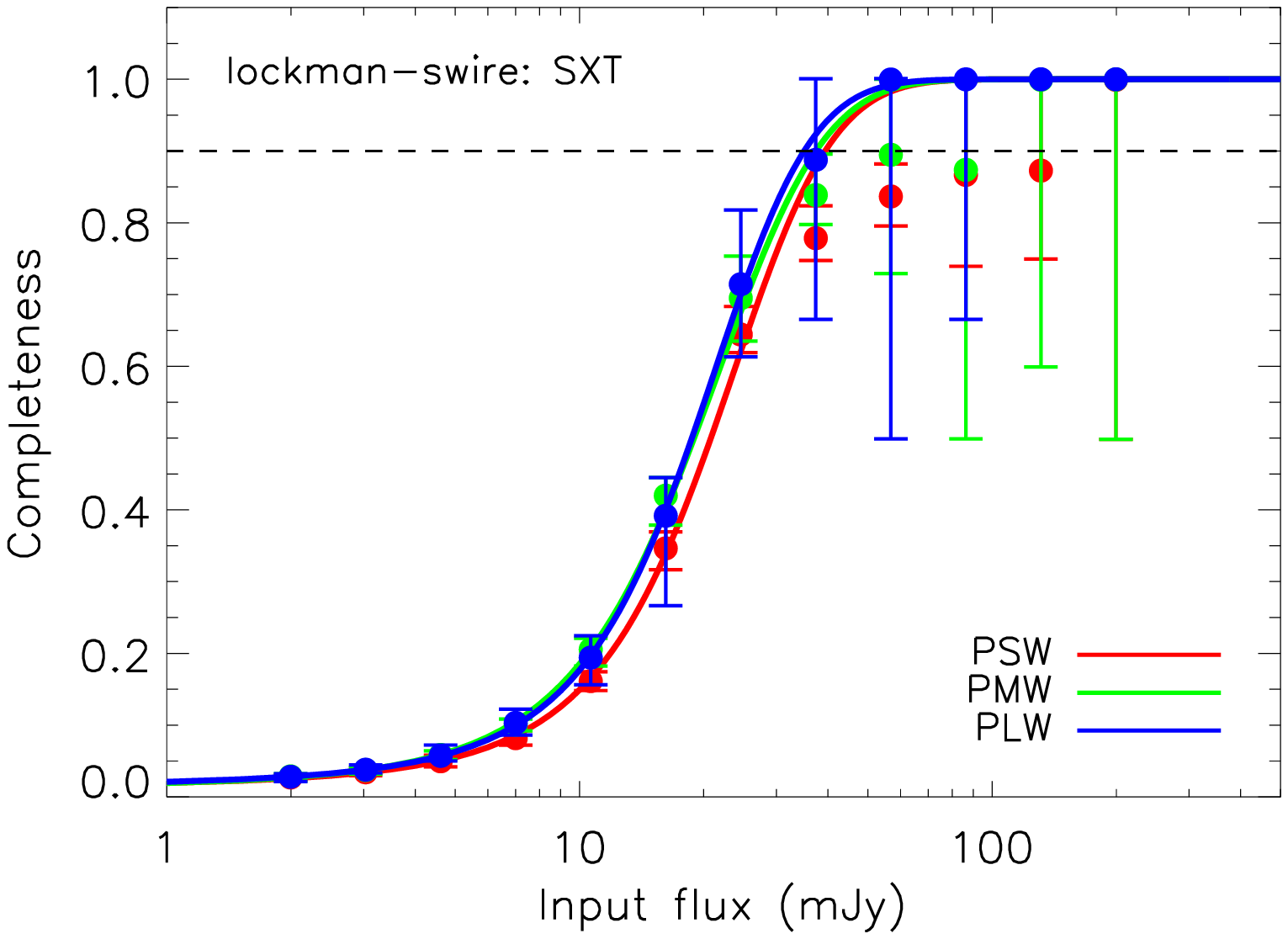}
\includegraphics[height=2.3in, width=3.3in]{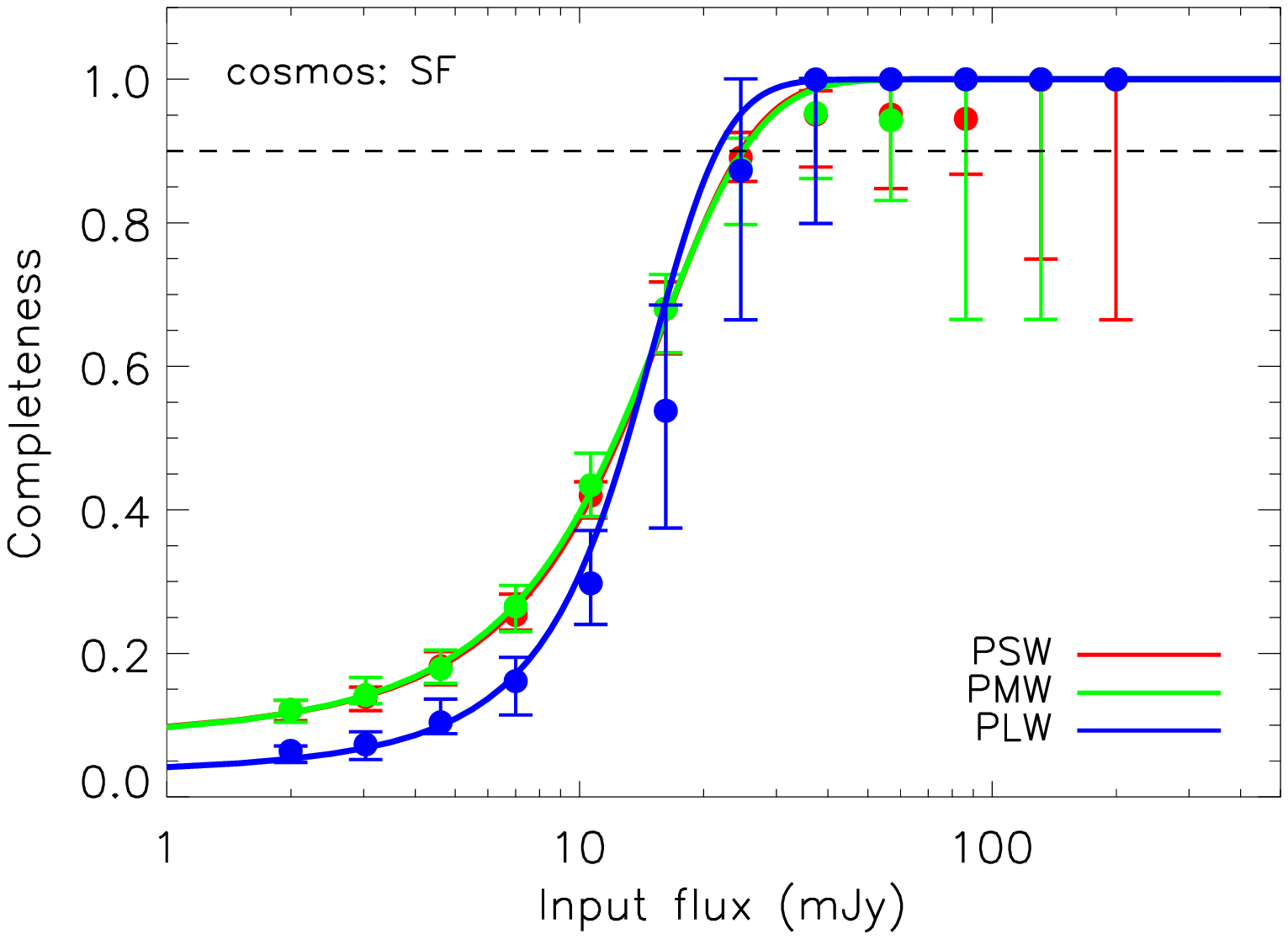}
\includegraphics[height=2.3in, width=3.3in]{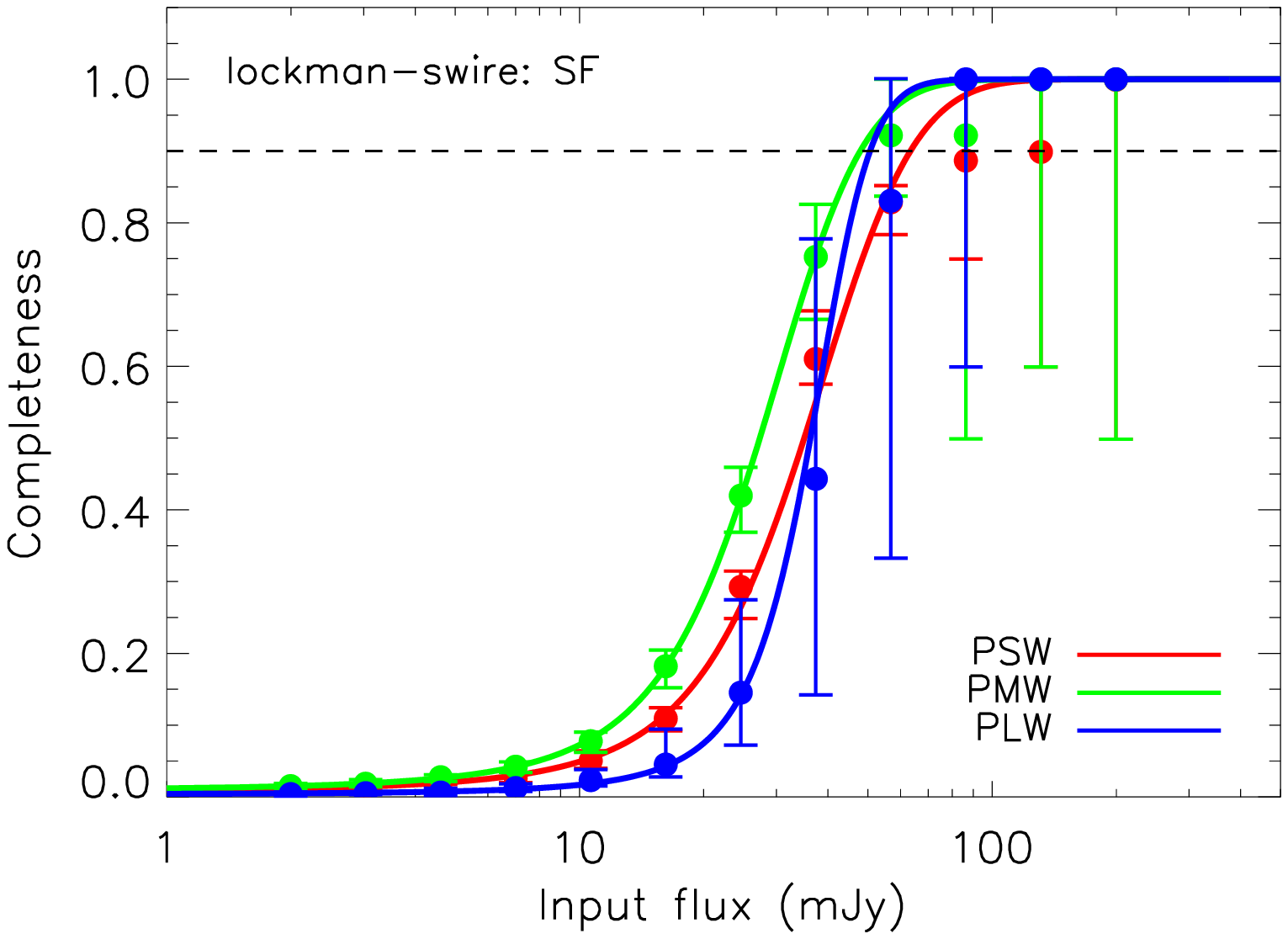}
\includegraphics[height=2.3in, width=3.3in]{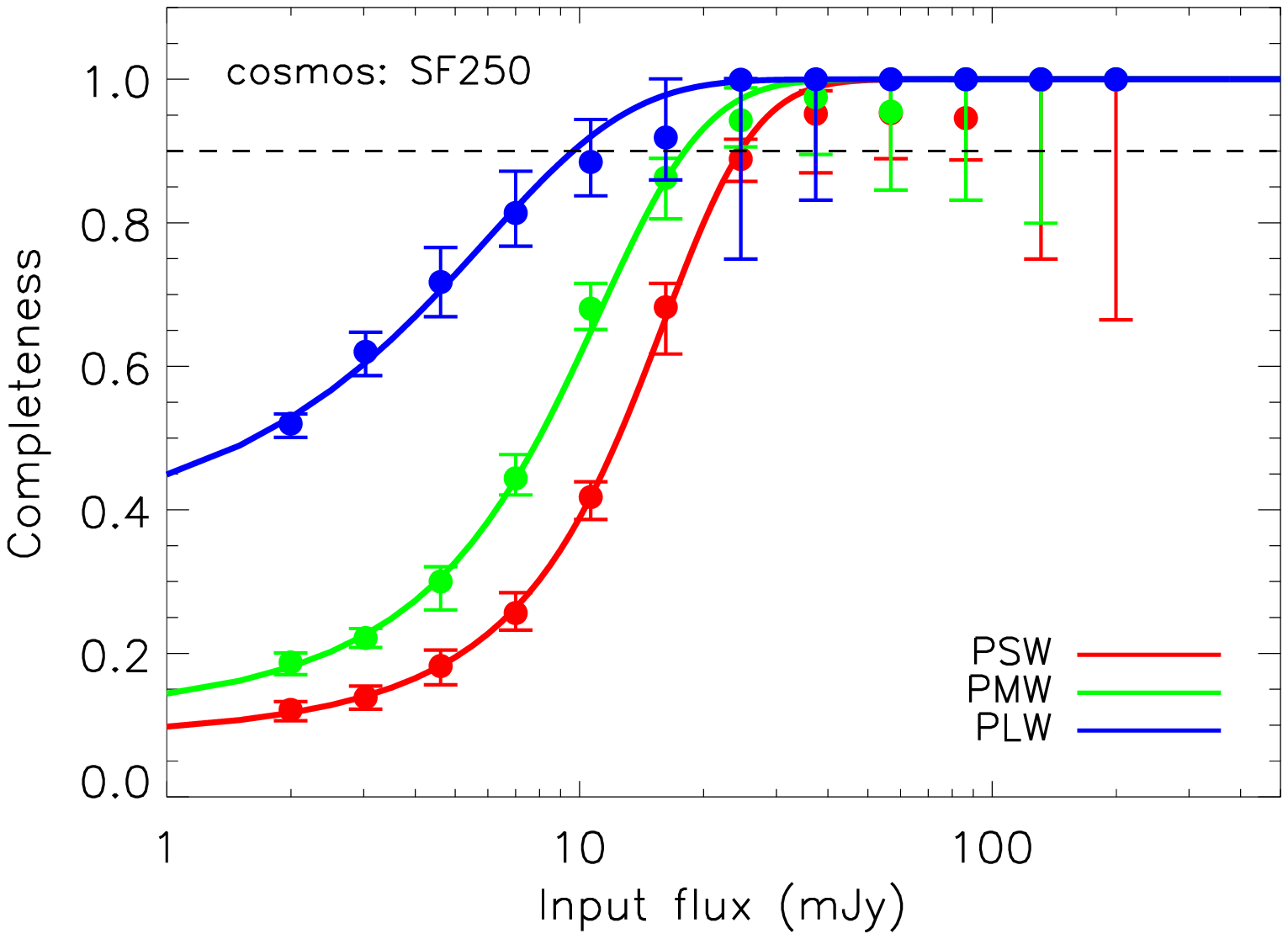}
\includegraphics[height=2.3in, width=3.3in]{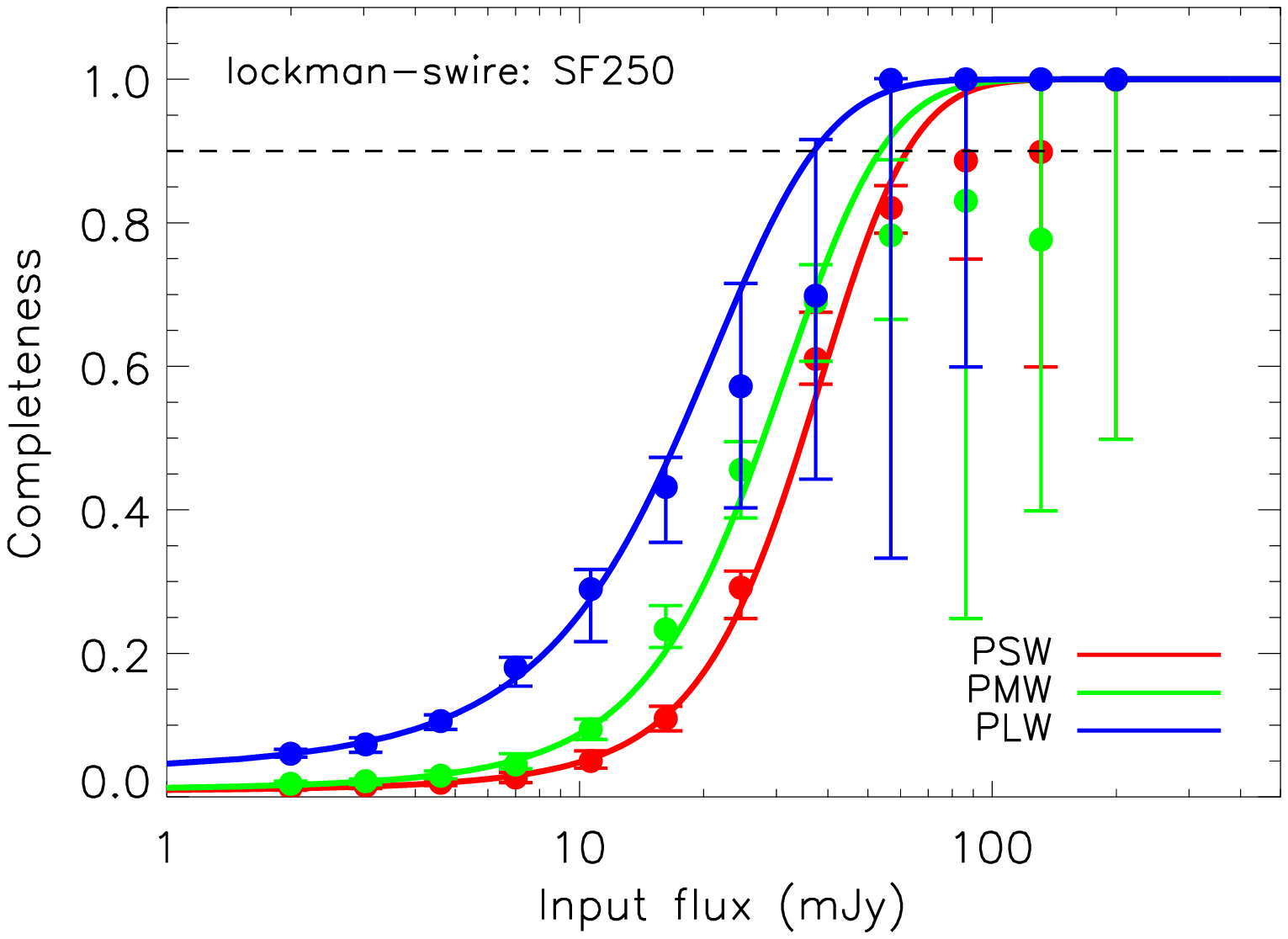}
\caption{The completeness fraction as a function of input flux density at 250, 350 and 500 $\micron$ for the three different types of source catalogues (top: SXT; middle: SF; bottom: SF250). The horizontal dashed line marks the $90\%$ completeness level. The left column corresponds to the simulated unclustered COSMOS field and the right column corresponds to the simulated unclustered Lockman-SWIRE field. The curves are the best-fit generalised logistic functions to describe the completeness ratio as a function of input flux density. Simulations with clustered input sources  produce  similar completeness curves.}
\label{comp}
\end{figure*}

\begin{table*}
\caption{The output-input flux difference to input flux ratio at 250, 350 and 500 $\micron$ for SXT, SF and SF250 catalogues extracted from unclustered simulations. For each catalogue, we give the interpolated output-input flux difference to input flux ratio at a given flux node based on the best-fit geometric function (Eq. 8) in the simulated COSMOS, EROTH, UDS, ELAIS-S1 and Lockman-SWIRE field. Clustered simulations give similar output-input flux difference to input flux ratios.}

\begin{tabular}{llll}
\hline
$S_{250}$ (mJy) & $\frac{S_{\rm out} - S_{\rm in}}{S_{\rm in}}$ (SXT; random) & $\frac{S_{\rm out} - S_{\rm in}}{S_{\rm in}}$ (SF; random)  &  \\
\hline
5 & 1.32/1.27/1.29/2.00/2.47 & 2.11/1.56/1.51/2.51/3.16& \\
10 & 0.39/0.43/0.44/0.71/0.90& 0.76/0.52/0.50/0.86/1.10&\\
20 & 0.11/0.15/0.15/0.25/0.33&0.27/0.18/0.17/0.30/0.38&\\
40 & 0.03/0.06/0.06/0.09/0.13 &0.10/0.06/0.06/0.10/0.13&\\
80 & 0.01/0.02/0.02/0.04/0.05&0.03/0.02/0.02/0.03/0.04&\\
160 & 0.00/0.01/0.01/0.02/0.02&0.01/0.01/0.01/0.01/0.01&\\
\hline
$S_{350}$ (mJy) & $\frac{S_{\rm out} - S_{\rm in}}{S_{\rm in}}$ (SXT; random) & $\frac{S_{\rm out} - S_{\rm in}}{S_{\rm in}}$ (SF; random)  & $\frac{S_{\rm out} - S_{\rm in}}{S_{\rm in}}$ (SF250; random) \\
\hline
5 &  1.31/1.20/1.28/1.64/1.98&1.93/1.34/1.30/2.03/2.50 & 1.11/0.76/0.74/1.31/1.52\\
10 & 0.40/0.41/0.48/0.60/0.73&0.69/0.45/0.45/0.72/0.91&0.35/0.21/0.20/0.41/0.48\\
20 & 0.12/0.14/0.18/0.22/0.27&0.25/0.15/0.15/0.25/0.33&0.11/0.05/0.05/0.12/0.15\\
40 & 0.03/0.05/0.07/0.08/0.10&0.09/0.05/0.05/0.09/0.12&0.03/0.01/0.01/0.04/0.05\\
80 &0.01/0.01/0.03/0.03/0.04 & 0.03/0.01/0.02/0.03/0.04&0.01/0.00/0.00/0.01/0.01\\
160 &0.00/0.00/0.02/0.01/0.01 &0.01/0.00/0.00/0.01/0.01&0.00/0.00/0.00/0.00/0.00\\
\hline      
$S_{500}$ (mJy) & $\frac{S_{\rm out} - S_{\rm in}}{S_{\rm in}}$ (SXT; random) & $\frac{S_{\rm out} - S_{\rm in}}{S_{\rm in}}$ (SF; random)  & $\frac{S_{\rm out} - S_{\rm in}}{S_{\rm in}}$ (SF250; random) \\
\hline
5 & 1.23/1.87/1.58/2.46/3.04 & 1.89/1.89/1.82/2.64/3.36& 0.21/0.20/0.29/0.51/0.58 \\
10 & 0.40/0.71/0.60/0.94/1.20&0.67/0.72/0.65/0.99/1.32&0.02/0.04/0.07/0.14/0.15\\
20 & 0.13/0.27/0.23/0.36/0.47&0.23/0.26/0.23/0.37/0.51&-0.0/0.00/0.02/0.04/0.04\\
40 & 0.04/0.10/0.09/0.13/0.18&0.08/0.09/0.08/0.13/0.20&-0.0/-0.0/0.00/0.01/0.00\\
80 & 0.01/0.04/0.03/0.05/0.06&0.02/0.02/0.03/0.05/0.07& -0.0/-0.0/0.00/0.00/0.00\\
160 & 0.00/0.01/0.01/0.01/0.02&0.00/-0.0/0.01/0.01/0.02&-0.0/-0.0/0.00/0.00/-0.0\\
\hline
\end{tabular}
\label{tab:flux_prob}
\end{table*}

Having determined all the necessary positional and photometric PDFs, we can now calculate the LR of all matches between the input and output catalogue. But we still need to isolate the real matches between the input and output from the random matches. When the noise in the data is entirely due to instrumental effects, the probability that a detection is genuine (or spurious) can be estimated from the SNR of the source. However, in these \textit{Herschel}--SPIRE data, the dominant source of noise is in general the confusion noise. So, the measurement of the flux density of any particular source is contaminated by the flux density of neighbouring sources. This means that the signal-to-(total) noise of a detection can not be used in a straightforward way to give the probability that it is spurious. To circumvent this problem, we match the randomised output catalogue with the input catalogue and calculate the LR of each matched pair, which basically characterises the LR distribution of spurious matches between the input and the output. As a result, we can derive the false identification rate\footnote{The false identification rate is defined as the ratio of the number of matches between the input and randomised output catalogue above a chosen LR threshold to the total number of matches.} as a function of LR threshold.  Finally, we select all matches between the input catalogue and the output catalogue with LR above the $10\%$ false identification rate as the true input-output matches.

\begin{table*}
\caption{The completeness fraction at 250, 350 and 500 $\micron$ for SXT, SF and SF250 catalogues extracted from unclustered simulations. For each catalogue, we give the interpolated completeness level at a given flux node based on the best-fit generalised logistic function (Eq. 9) in the simulated COSMOS, EROTH, UDS, ELAIS-S1 and Lockman-SWIRE field. Clustered simulations give similar completeness fractions.}
\begin{tabular}{llll}
\hline
$S_{250}$ (mJy) & Comp (SXT; random) & Comp (SF; random)  &  \\
\hline
5 &  0.14/0.11/0.12/0.08/0.06 &0.19/0.13/0.13/0.04/0.02 & \\
10 & 0.29/0.28/0.31/0.21/0.15&0.39/0.32/0.34/0.09/0.05 &\\
20 & 0.66/0.66/0.71/0.59/0.47&0.80/0.81/0.79/0.36/0.17&\\
40 & 0.97/0.96/0.97/0.95/0.91&0.99/1.00/0.99/0.91/0.60&\\
80 & 1.00/1.00/1.00/1.00/1.00&1.00/1.00/1.00/1.00/0.97&\\
160 & 1.00/1.00/1.00/1.00/1.00& 1.00/1.00/1.00/1.00/1.00&\\
\hline
$S_{350}$ (mJy) & Comp (SXT; random) & Comp (SF; random)  & Comp (SF250; random) \\
\hline
5 &0.15/0.09/0.10/0.08/0.07&0.20/0.15/0.15/0.05/0.03& 0.33/0.21/0.22/0.07/0.03\\
10 &0.31/0.22/0.25/0.22/0.18& 0.40/0.37/0.38/0.13/0.07& 0.61/0.49/0.51/0.18/0.09\\
20 &0.66/0.68/0.61/0.59/0.54&0.80/0.82/0.79/0.46/0.28&0.93/0.90/0.88/0.56/0.30\\
40 & 0.96/0.99/0.94/0.94/0.92&0.99/0.99/0.99/0.93/0.80&1.00/1.00/1.00/0.96/0.75\\
80 & 1.00/1.00/1.00/1.00/1.00&1.00/1.00/1.00/1.00/1.00&1.00/1.00/1.00/1.00/0.99\\
160 &1.00/1.00/1.00/1.00/1.00&1.00/1.00/1.00/1.00/1.00&1.00/1.00/1.00/1.00/1.00\\
\hline
$S_{500}$ (mJy) & Comp (SXT; random) & Comp (SF; random)  & Comp (SF250; random) \\
\hline
5 & 0.25/0.16/0.15/0.09/0.06& 0.11/0.07/0.09/0.02/0.01&0.73/0.49/0.52/0.23/0.12\\
10 &0.52/0.45/0.39/0.26/0.18&0.31/0.28/0.25/0.05/0.02& 0.91/0.78/0.76/0.45/0.25\\
20 & 0.89/0.94/0.84/0.73/0.55&0.86/0.90/0.75/0.29/0.07&  0.99/0.98/0.96/0.79/0.58\\
40 &1.00/1.00/1.00/0.99/0.94&1.00/1.00/0.99/0.97/0.64&1.00/1.00/1.00/0.98/0.92\\
80 & 1.00/1.00/1.00/1.00/1.00&1.00/1.00/1.00/1.00/1.00&1.00/1.00/1.00/1.00/1.00\\
160 &1.00/1.00/1.00/1.00/1.00&1.00/1.00/1.00/1.00/1.00&1.00/1.00/1.00/1.00/1.00\\
\hline
\end{tabular}
\label{tab:flux_prob}
\end{table*}

\subsection{Photometric accuracy and completeness}

Having matched the input and output catalogue, we can look at the photometric accuracy of the extracted sources. In Fig.~\ref{fluxunc}, we plot the output - input flux difference ($S_{\rm out} - S_{\rm in}$) to input flux ($S_{\rm in}$) ratio for extracted sources in different bins of input flux density at 250, 350 and 500 $\micron$, for the three different types of source catalogues (SXT, SF and SF250) in the simulated unclustered COSMOS and Lockman-SWIRE field. Simulations with clustered input sources give similar results. At the faint end ($<5\sigma$ limit), the output flux is generally larger than the input flux (the well-known flux-boosting effect) and the level of flux-boosting increases with decreasing input flux. At the bright end (flux densities $>5\sigma$ limit), the mean flux difference to input flux ratio $(S_{\rm out} - S_{\rm in})/S_{\rm in} $ stays close to zero (the dashed line in Fig.~\ref{fluxunc}) with an increasing scatter towards deceasing input flux. For each of the three types of source catalogues, the ratio $(S_{\rm out} - S_{\rm in})/S_{\rm in} $ deviates from the dashed line at a larger $S_{\rm in}$ value in the simulated Lockman-SWIRE field than the COSMOS field, due to a higher level of instrument noise in the former (see Table 2). For the band-merged SF catalogues (SF250) extracted at the positions of SF 250 $\micron$ sources, the ratio $(S_{\rm out} - S_{\rm in})/S_{\rm in} $ deviates from the dashed line at a much smaller $S_{\rm in}$ value at 350 and 500 $\micron$ compared to the independent  single-band SXT or SF catalogues, as a result of reduced confusion noise. We fit a geometric function of the form 
\begin{equation}
\left (\frac{S_{\rm out} - S_{\rm in}}{S_{\rm in}} \right) = a_0 (S_{\rm in})^{a_1} + a_2
\end{equation} 
to describe the relation between the mean flux difference to input flux ratio as a function of input flux density. Here $a_1$ is negative. So, as the input flux $S_{\rm in}$ increases to a very large number, the flux difference to input flux ratio asymptotes to $a_2$, $(S_{\rm out} - S_{\rm in})/S_{\rm in} =a_2$. $a_1$ describes how quickly the flux difference to input flux ratio rises (i.e. deviates from the asymptotic value $a_2$) as a function of decreasing $S_{\rm in}$ and $a_0$ is related to the input flux density at which $(S_{\rm out} - S_{\rm in})/S_{\rm in}$ starts to deviate from  $a_2$. In other words, $a_1$ describes the rate of deviation and $a_0$ is related to the deviation point. In Table 4, we list the best-fit values for the parameters in the geometric function at 250, 350 and 500 $\micron$ averaged over all five simulated fields, both clustered and unclustered. In all cases, $a_2$ is consistent with zero which means for the bright input sources there is no systematic bias in the flux estimation in the SXT, SF and SF250 catalogues. We can see that averaged over different fields the rate of deviation $a_2$ is similar across different bands for the SXT and SF catalogues. The SF250 catalogues have a higher rate of deviation (i.e. steeper rise) and lower deviation point compared to the SXT and SF catalogues at 350 and 500 $\micron$. Simulations with clustered input sources give in general similar results to the unclustered simulations but with a slightly higher deviation point. In Table 5, we list the interpolated output-input flux difference to input flux ratio at 250, 350 and 500 $\micron$ as a function of input flux based on the best-fit geometric function in the unclustered simulation of the COSMOS, EROTH, UDS, ELAIS-S1 and Lockman-SWIRE field. Clustered simulations give similar output-input flux difference to input flux ratio. 

Completeness fraction is defined as the ratio of the number of input sources matched with sources in the output catalogue (i.e. detected input sources) to the total number of input sources in a given flux interval. As such, completeness fraction is defined as a function of input flux. After matching the input catalogue with the output catalogue as detailed in Section 3.2, it is straightforward to derive the completeness curve for each of our simulations. Input sources that are linked to more than one output source are counted only once to avoid double counting. Fig. ~\ref{comp} compares the completeness curves from SXT,  SF and SF250 catalogues at 250, 350 and 500 $\micron$ in the simulated unclustered COSMOS and Lockman-SWIRE field. In deep fields, SF catalogues are deeper than the SXT fields. In shallower fields, the opposite is true. A higher level of instrument noise in the shallow fields means that fewer sources would pass the correlation test between the source profile and the PRF in the SF source detection method. For a given source extraction method (SXT or SF), the completeness fraction at a given flux density drops as the level of instrument noise increases. Simulations with clustered sources produce similar completeness curves as a function of input flux.
To fit the completeness fraction as a function of input flux density, we can use the generalised logistic function\footnote{The (generalised) logistic function is a type of sigmoid function (``S''-shaped function), often used to model population growth. It shows initial exponential growth when the independent variable is small, followed by slower growth with increasing values of the independent variable, and eventually reaches saturation point when the independent variable is large.}
\begin{equation}
C(S_{\rm in}) = A +\frac{K-A}{(1 + Q\exp{(-(S_{\rm in}-M)/B)})^{N }}. 
\end{equation}
We can set $A=0$ (the lower asymptote) and $K=1$ (the upper asymptote)  as the completeness fraction $C$ should approach 100\% and 0\% when $S_{\rm in}$ goes to very large and very small values respectively. In Table 6, we list the interpolated completeness level at 250, 350 and 500 $\micron$ as a function of input flux based on the best-fit generalised logistic function in the unclustered simulation of  the COSMOS, EROTH, UDS, ELAIS-S1 and Lockman-SWIRE field. Clustered simulations give similar results. 

\section{Extended sources}

\begin{figure}
    \centering
        \includegraphics[height=1.in, width=1.in]{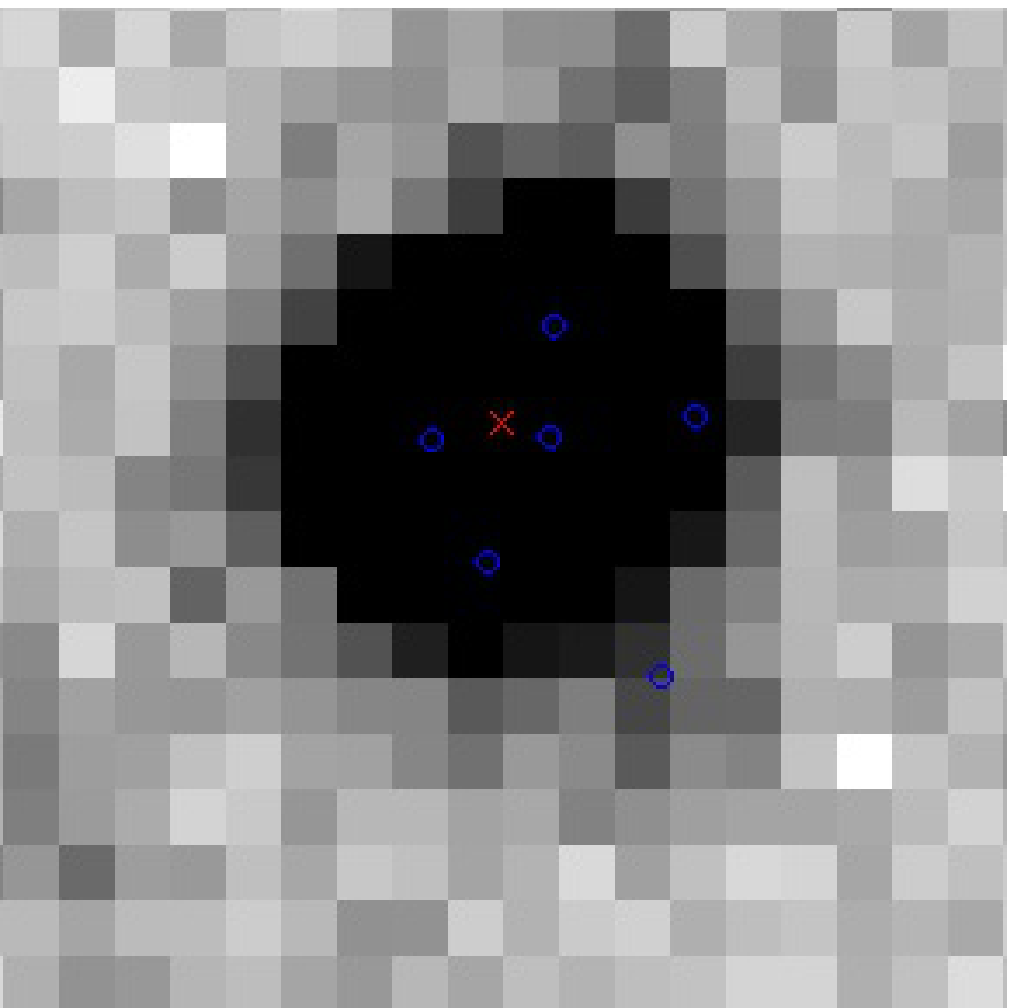}
        \includegraphics[height=1.in, width=1.in]{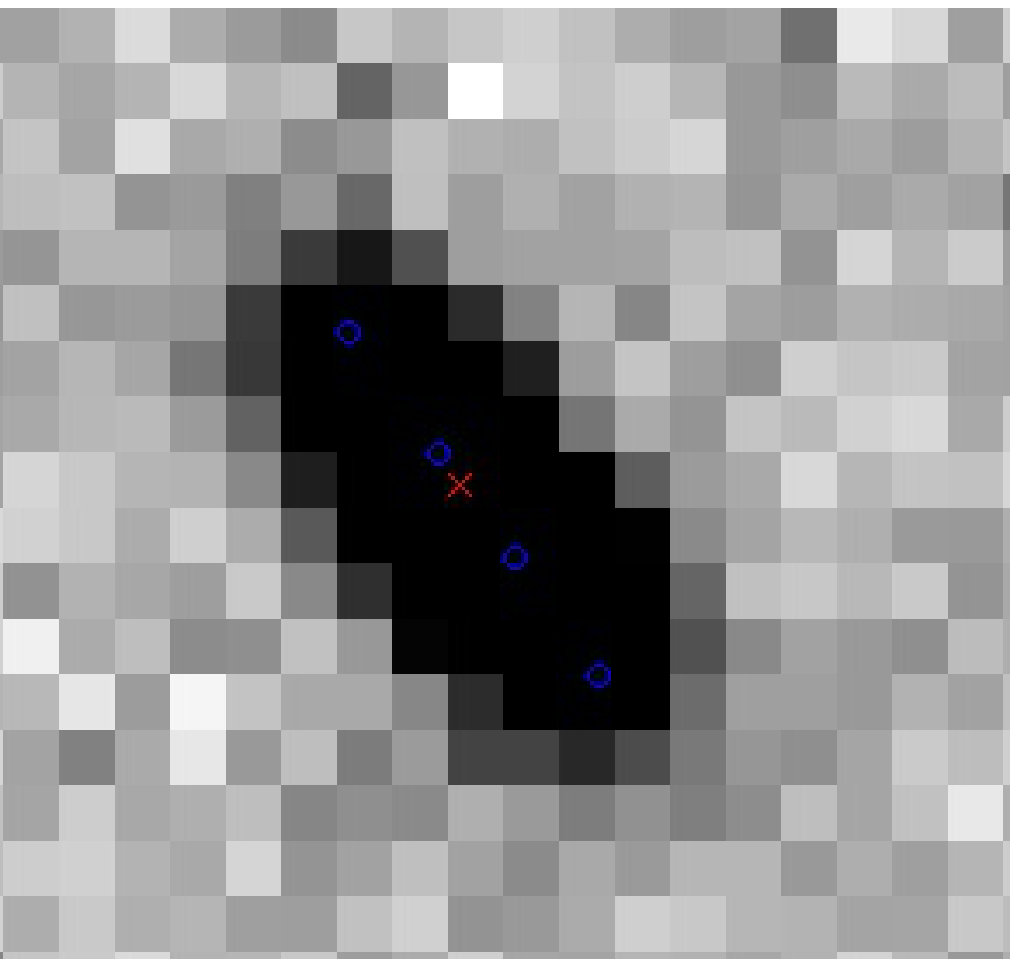}
        \includegraphics[height=1.in, width=1.in]{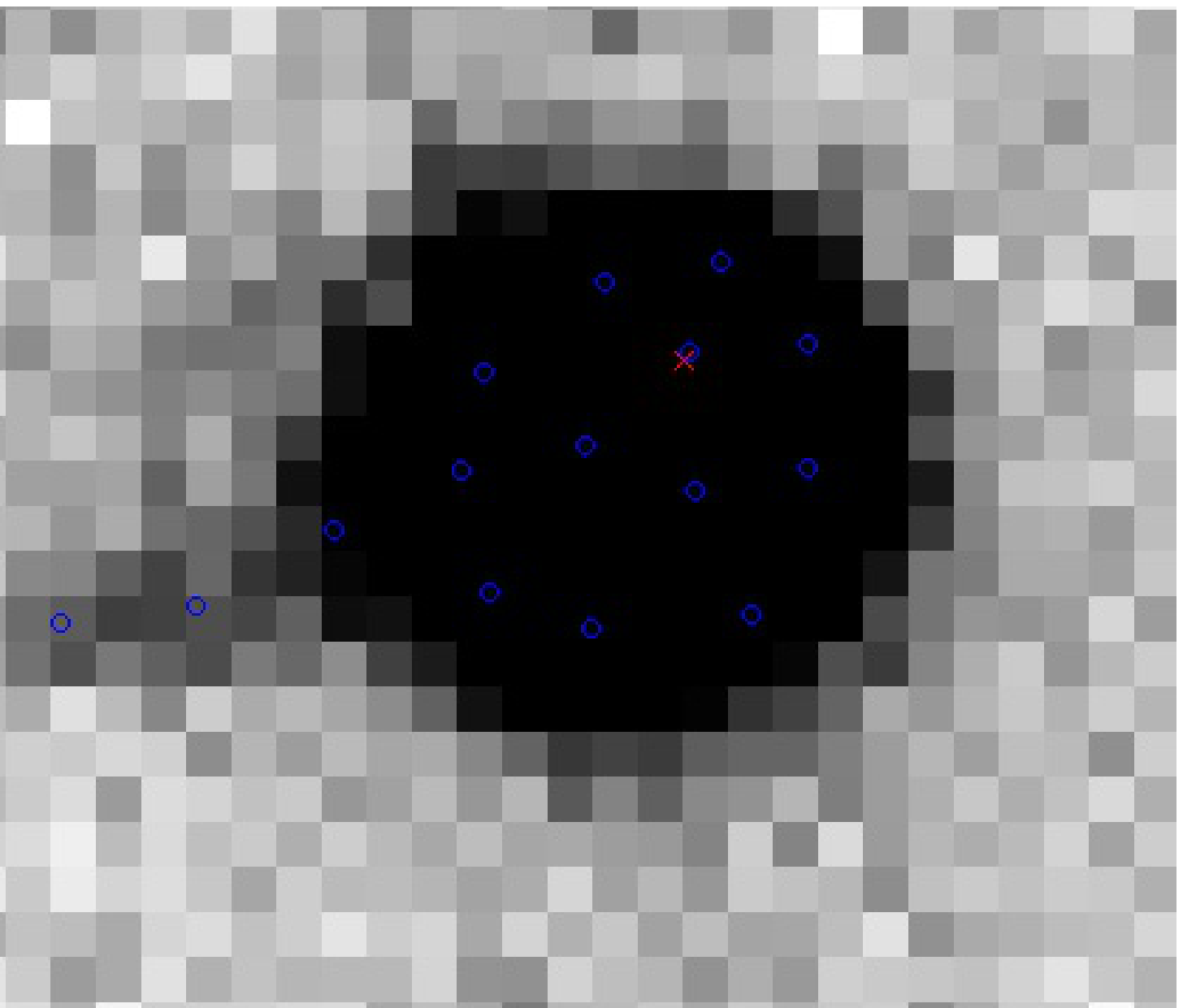}
    \caption{Examples of extended sources in the lockman-swire region.} 
    \label{extendedsources}
\end{figure}

Our source detection and extraction methods assume point sources.  Objects that are extended on the scale of the SPIRE beam (see Fig.~\ref{extendedsources}) are not expected to be accurately represented.  The SXT detection and extraction will lead to inaccurate flux estimation (probably biased low) and very large objects may be misidentified as multiple point sources.  Similarly, SF will underestimate fluxes and is expected to break even modestly extended object into multiple point sources.  We have thus flagged detections as extended using the following criteria.  


We have cross-matched all SPIRE detections with their nearest (within 30\arcsec) counterpart in the 2MASS Extended Source Catalog (XSC, Jarrett et al. 2000). Any detection matched to a counterpart with a K fiducial Kron elliptical aperture semi-major axis $>$ 9\arcsec\ is flagged as extended.  We have then cross-matched all SPIRE detections with sources in the new catalogue of principal galaxies (PGC2003) which constitutes the HyperLeda database\footnote{PGC2003 contains about one million confirmed galaxies flux limited to $\sim18$ $B$-mag. HyperLeda (\url{http://leda.univ-lyon1.fr}) provides the richest catalogue of homogeneous parameters of galaxies for the largest available sample, and is thus useful when trying to estimate galaxy sizes in an homogeneous manner.} (Paturel et al, 2003). A counterpart is identified where a 30\arcsec\ radius circle around the SPIRE position intersects with the 25 $B$-mag/arcsec$^2$ isophotal ellipse. Any detection with a counterpart with diameter at this isophot (D25) $>$ 18\arcsec\  is flagged as extended.  These scales were chosen as they correspond to the FWHM of the SPIRE beam at 250 $\mu$m and could conservatively reproduce other classifications, i.e. all objects which we have identified as bright and extended or nearby by eye and clusters of four or more SF detections.  This procedure flags 6,169 SPIRE detections as extended. The number of detections is much larger than the number of extended galaxies they represent as the detections include objects detected by either of the two techniques, in any SPIRE band, and multiple components of a single galaxy.

Another test which informed and supplemented our flagging method was an analysis of SPIRE detected sources with a poor PRF fit at the position of the Spitzer 24 $\mu$m position. Visual inspection revealed those detections appeared to be extended. Most of these ($\sim$90\%) have already been flagged as extended by the methods described above.  The remaining 10\% (200 detections) were then also flagged as extended which brings the total number of extended sources to 6,369.  However, this additional safety check has only been done in regions with Spitzer 24 $\mu$m data.


\section{Conclusions and Discussions}

In this paper, we present independent single-band SUSSEXtractor (SXT) and StarFinder (SF) point source catalogues as well as band-merged SF catalogues (SF250) extracted at the positions of the SF 250 $\mu$m sources released in HerMES DR1 and DR2. For SF and SF250 catalogues, we use our own code DESPHOT for accurate photometry. End-to-end simulations with realistic number counts and clustering behaviour matched to the observed counts and power spectra of the SPIRE sources are generated to characterise the basic properties of the source catalogues. 

We use a likelihood ratio method to match the simulated input sources with the output sources. The matched input and output catalogues are estimated to have a false identification rate of $10\%$. We find that the positional distribution of real matches between the input and output peaks at approximately 5\arcsec, 8\arcsec and 13\arcsec\ at 250, 350 and 500 $\micron$ for SXT catalogues, and approximately at 5\arcsec, 7\arcsec and 12\arcsec\ at 250, 350 and 500 $\micron$ for SF catalogues. Both source extraction methods (SXT and SF) return unbiased flux measurement for bright sources at $>5\sigma$ (with respect to the total noise including confusion noise and instrument noise). For faint sources, the output flux systematically overestimates the input flux and the level of flux-boosting (the output-input flux difference to input flux ratio) increases rapidly with decreasing input flux. At a given input flux, the level of flux boosting also increases with the level of instrument noise. The completeness fraction as function of input flux is also characterised for the three different types of source catalogues based on our simulations. In the deep fields, SF catalogues are generally deeper than the SXT catalogues. In the shallower fields with a higher instrument noise level, the opposite is true. By construction, the SF250 catalogues are deeper than the independent SF catalogues at 350 and 500 $\micron$ but it will miss sources which are only detected at 350 and/or 500 $\micron$. Fitting formulae for the positional accuracy, photometric accuracy and completeness fraction are given in the paper. We find that the impact of source clustering on the positional and photometric accuracy as well as the completeness fraction is small.

\section*{ACKNOWLEDGEMENTS}

LW is supported by UK's Science and Technology Facilities Council grant ST/F002858/1 and an ERC StG grant (DEGAS-259586). The data presented in this paper will be released through the Herschel database in Marseille HeDaM (hedam.oamp.fr/herMES). SPIRE has been developed by a consortium of institutes led by Cardiff Univ. (UK) and including Univ. Lethbridge (Canada); NAOC (China); CEA, LAM (France); IFSI, Univ. Padua (Italy); IAC (Spain); Stockholm Observatory (Sweden); Imperial College London, RAL, UCL-MSSL, UKATC, Univ. Sussex (UK); Caltech, JPL, NHSC, Univ. Colorado (USA). This development has been supported by national funding agencies: CSA (Canada); NAOC (China); CEA, CNES, CNRS (France); ASI (Italy); MCINN (Spain); SNSB (Sweden); STFC (UK); and NASA (USA).

\end{document}